\documentclass{aa}  
\usepackage{graphicx}
\usepackage{epstopdf}
\usepackage{lscape} 
\usepackage{txfonts}
\usepackage{natbib} 
\bibpunct{(}{)}{;}{a}{}{,} 
\begin{document}
\title{Point spread functions for the Solar Optical Telescope onboard Hinode}
\titlerunning{Point spread functions for SOT}

\author{Sven Wedemeyer-B\"ohm\inst{1,2}\thanks{Marie Curie Intra-European Fellow of the European Commission}}

\offprints{sven.wedemeyer-bohm@astro.uio.no}

\institute{Institute of Theoretical Astrophysics, University of Oslo,
  P.O. Box 1029 Blindern, N-0315 Oslo, Norway
  \and
  Center of Mathematics for Applications (CMA), University of Oslo,
  Box 1053 Blindern, NÐ0316 Oslo, Norway
}

\date{Received 20 March 2008; accepted 1 May 2008}

\abstract{}
{The combined point spread function (PSF) of the  Broadband Filter Imager  
(BFI) and the Solar Optical Telescope (SOT) onboard the Hinode spacecraft  
is investigated.  
}
{Observations of the Mercury transit from November 2006 and the solar eclipse(s) 
from 2007 are used to determine the PSFs of SOT for the blue, green, and red continuum 
channels of the BFI.  
For each channel large grids of theoretical point spread functions are calculated
by convolution of the ideal diffraction-limited PSF and Voigt profiles.  
These PSFs are applied to artificial images of an eclipse and a Mercury transit. 
The comparison of the resulting artificial intensity profiles across the terminator 
and the corresponding observed profiles yields a quality measure for each case. 
The optimum PSF for each observed image is indicated by the best fit. 
} 
{The observed images of the Mercury transit and the eclipses exhibit a clear 
proportional relation between the residual intensity and the overall light level 
in the telescope. 
In addition there is a anisotropic stray-light contribution.  
These two factors make it very difficult to pin down a single unique PSF that can 
account for all observational conditions. 
Nevertheless the range of possible PSF models can be limited by using  
additional constraints like the pre-flight measurements of the Strehl ratio.
} 
{BFI/SOT operate close to the diffraction limit and have only a rather 
small stray-light contribution. 
The FWHM of the PSF is broadened by only $\sim 1$\,\% with respect to the 
diffraction-limited case, while the overall Strehl ratio is $\sim 0.8$. 
In view of the large variations -- best seen in the residual intensities of 
eclipse images -- and the dependence on the overall light level and position in the 
FOV, a range of PSFs should be considered instead of a single PSF per wavelength. 
The individual PSFs of that range allow then the determination of 
error margins for the quantity under investigation. 
Nevertheless the stray-light contributions are here found to be best 
matched with Voigt functions with the parameters $\sigma = 0\,\farcs008$ and 
$\gamma = 0\,\farcs004$, 0\,\farcs005,  and 0\,\farcs006 for the blue, green, 
and red continuum channels, respectively.
} 


\maketitle
%
\section{Introduction}
\label{sec:intro}

The quantitative interpretation of an observation requires a good knowledge of 
the properties of the utilised instrument as it can significantly affect 
empirically derived properties. 
A realistic model of the instrumental influence is of particular importance for 
the comparison of certain aspects of observations and numerical models, e.g. 
for the intensity contrast of solar granulation
\citep{1984ssdp.conf..174N}.

Each optical instrument causes the radiation emitted by a point source (or an 
object) to be spread on the image plane of the detector. 
The image of an extended light source like the Sun is degraded this way 
and small spatial scales remain unresolved. 
Mathematically this phenomenon is described by means of a point spread 
function (PSF) or its Fourier transform, the Optical Transfer Function (OTF).  
Even an otherwise perfect optical telescope induces image degradation 
due to diffraction at the finite aperture stop. 
The resulting diffraction-limit effectively defines the best theoretically 
achievable angular resolution of a telescope. 

In reality there is additional degradation due to (i)~the Earth's atmosphere 
(for ground-based instruments) and (ii)~instrumental effects. 
The terrestrial atmosphere introduces image degradation due to turbulence 
along the line of sight -- the so-called {\em seeing} -- and scattering. 
Seeing has mostly an impact on the central parts of the PSF, blurring the 
image, but also contributes to the wings, whereas scattering becomes 
significant for the far wings of the PSF \citep{1983SoPh...87..187M}. 
It depends on the elevation of the line of sight and weather conditions 
(e.g. dust, aerosols or water vapour in the atmosphere). 

Space-borne instruments have the advantage of not being limited by the Earth's 
atmosphere but still  instrumental stray-light has to be considered. 
Often the stray-light contribution is anisotropic and varies over the 
field of view (FOV), making it hard to derive a detailed model for the PSF. 
This effect is already visible in solar limb observations, where the intensity 
does not decrease to zero but to a finite value\footnote{The residual intensity outside the 
solar disc, caused by stray-light, is often called \textit{aureole}.}.
The instrumental deviations from the diffraction-limit are mostly caused 
by imperfections and contamination of the optical components in the telescope, 
e.g. dust on optical surfaces or impurity of lens material but also  microscopic 
scratches or micro-roughness. 
Also subtle effects like jitter of the spacecraft can contribute to the PSF.
The resulting (anisotropic) stray-light increases the relative intensity 
contribution from the PSF wings and with it the relative height of the 
side-lobes, which nevertheless stay in position. 
With a higher contribution from the wings the relative height of the central peak
and with it the corresponding 
Strehl ratio\footnote{The {\em Strehl (intensity) ratio}
is defined as the ratio between the central peak values of a PSF and the 
one of an ideal, i.e. aberration-free, diffraction-limited PSF. 
It provides a measure for the deviation from the diffraction limit.}
decrease, while the central peak is broadened.

Constructing a realistic PSF is unfortunately a non-trivial task 
as often important contributions are hardly known. 
This is in particular true for stray-light produced inside an optical instrument 
\citep{1983SoPh...87..187M}.
The best way to determine the properties of a telescope is of course 
the direct measurement in the laboratory but for space-borne instruments 
it is usually difficult to produce the same conditions as for the later 
operation.  
Another possibility of measuring the PSF is the observation of a partial solar eclipse. 
An illustrative example is the work by 
\citet{1975A&A....45..167D} who used observations of a partial eclipse in 1973. 
They fitted the observed intensity profile across the eclipse terminator with 
a combination of two Gaussians \citep[cf.][]{1971A&A....14...15L}. 
\cite{1984ssdp.conf..174N} argues that the wings of the spread function give rise to 
uncertainties, which make it problematic to retrieve a definite analytical fit of the 
measured profile. 
He showed that instead of two Gaussian also a combination of two Lorentzians produces 
a fit of comparable quality. 
The related uncertainties can have a fundamental impact on the derived intensity 
contrast. 
\citeauthor{1984ssdp.conf..174N} proposes a correction factor of 1.5 with respect to the results of 
\citeauthor{1975A&A....45..167D}. 
Beside these uncertainties, a fundamental problem remains as the fit of an observed 
terminator profile usually assumes a symmetric PSF/OTF but hardly can account for  
asymmetric contributions. 
As we shall see in the case of SOT a real PSF can have significant anisotropic components.

In many cases, however, suitable observations of an eclipse or a transit of 
Mercury or Venus \citep[see, e.g.,][]{deForest_TRACE_PSF}
are not available. 
Instead one has to estimate PSF properties from the spatial 
power spectral density and other ad hoc arguments. 
Another possibility is to estimate the PSF by comparing the intensity distribution 
(histograms) of a particular observation with numerical models 
\citep[see., e.g.,][]{2003ApJ...597L.173S, leenaarts05, 2007ApJ...655..615L}. 
Often the PSF is approximated with a linear combination of an Airy and 
a Lorentz function, characterised by the two parameters $\alpha$, the height 
of the Lorentzian relative to the height of the Airy core, and $\beta$, the FWHM 
of the Lorentzian wings. 
In most cases, however, there is an extended range of possible combinations of 
$\alpha$ and~$\beta$. 
More severe is the (implicit) assumption that the numerical model, which is used as 
reference for the fit between observed and synthetic image, is perfectly realistic.
 
Here a set of PSFs for the Solar Optical Telescope 
\citep[SOT,][]{tsuneta_2008_SOToverview,2008arXiv0804.3248I, 
2008SoPh..tmp...26S,2007SoPh..tmp..154S}
onboard the Hinode spacecraft \citep{2007SoPh..243....3K} is derived 
from observations of a Mercury transit and two eclipses for different wavelengths. 
The observations are described and analysed in Sect.~\ref{sec:obs}. 
Theoretical two-dimensional PSFs are constructed by convolution of ideal 
diffraction-limited PSFs and Voigt functions and applied to an artificial 
eclipse and a Mercury transit in Sect.~\ref{sec:psfmodel}.
By comparison of the resulting intensity profiles and the observational results, 
the best PSF is determined for each case.
Discussion and conclusions follow in Sects.~\ref{sec:discus} and \ref{sec:conc}, 
respectively. 
SOT observations of quiet Sun granulation will be systematically analysed and 
compared to synthetic intensity images from numerical simulations 
in forthcoming papers.

\section{Observations}
\label{sec:obs}

\begin{figure*}[phpt]
   \centering
   \includegraphics[width=0.75\textwidth]{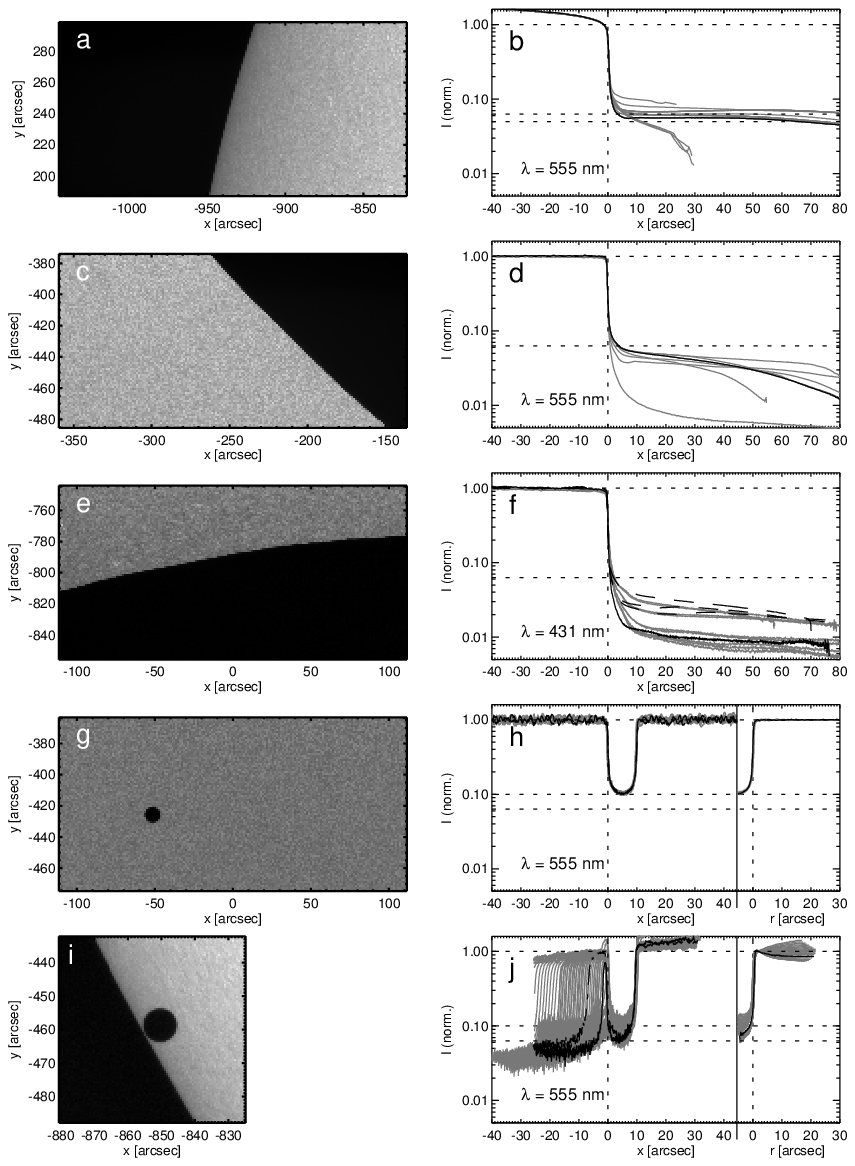}
   \caption{Hinode/SOT observations from top to bottom: 
   \textbf{a)} limb observation, 
   \textbf{c)} the total eclipse from March 19th, 2007, 
   \textbf{e)} the partial eclipse from February 17th, 2007, and 
   \textbf{g)} the Mercury transit  from November 8th 2006 on the solar disc 
   ($x = -52\,\arcsec$, $y = -426\,\arcsec$,  $\mu = 0.9$) and 
   \textbf{i)} at the solar limb.
   For all cases an exemplary image for a wavelength of 555\,nm is selected, 
   except for the partial eclipse which was only observed in G-band (431\,nm).
   The panels to the right show intensity profiles of the image shown on the left 
   (black solid) and for all other available images    
   of the same kind (dark-grey). 
   The profiles are averaged over all rows, aligned perpendicular to the 
   terminator/limb. The intensity is normalised to the intensity of the bright 
   solar disc close to the limb / terminator. 
   The left half of the panels h and j, show averages over the three rows 
   and columns through the centre of the Mercury disc, whereas the radial averages 
   can be seen in the right half. 
   The dotted lines are for orientation only. 
   For comparison profiles from the total eclipse observed in G-band are plotted 
   in panel~f (dashed lines).}
   \label{fig:sotstray}
\end{figure*}

\subsection{Filtergrams} 
The observations analysed in this study were obtained with the Broadband 
Filter Imager (BFI) of the Solar Optical Telescope (SOT) onboard the Hinode 
satellite (see Sect.~\ref{sec:intro} for references).
Primarily the blue, green and red continuum channels of the BFI are considered. 
The central wavelengths of these filters are 450.45\,nm, 555.05\,nm, and 668.40\,nm, 
respectively. 
All three filters have a band width of 0.4\,nm. 
In addition a few images for the other three BFI filters are used. 
These filters are CN, Ca\,II\,H, and G-band with central wavelengths of 
388.35\,nm, 396.85\,nm, and 430.50\,nm, respectively.  
The correponding band widths are 0.7\,nm, 0.3\,nm, and 0.8\,nm.

Observations of the Mercury transit from November 8th, 2006, and the solar eclipse 
of March 19, 2007, are analysed (see Fig.~\ref{fig:sotstray} for examples).
The eclipse was partial on Earth but total for Hinode. 
It is referred to as total eclipse hereafter. 
There also occured a partial eclipse on February 17, 2007, which was observed by 
Hinode but which was not visible from the ground. 
It was observed with the G-band channel only 
but still provides useful information for this study.  
Additional observations of quiet Sun granulation for various positions 
on the disc and at the solar limb are used. 

\begin{table*}
  \caption{Residual intensities $I_\mathrm{res}$ in the observations  
  at different distances $r$ from the terminator or the 
  solar limb, respectively. See text for details.} 
  \label{tab:iresobs}
  \centering
  \begin{tabular}{l|c@{ }c|c@{ }c|c@{ }c|c@{ }c|c@{ }c|c@{ }c}
  \hline
  \hline
  band          &\multicolumn{2}{c}{CN}&
  				 \multicolumn{2}{c}{Ca II H}&
  				 \multicolumn{2}{c}{G band}&
				 \multicolumn{2}{c}{bc}&
		  		 \multicolumn{2}{c}{gc}&
				 \multicolumn{2}{c}{rc}\\
  $\lambda$ [nm]&\multicolumn{2}{c}{388.3}&
  			     \multicolumn{2}{c}{396.9}&
  			     \multicolumn{2}{c}{430.5}&
			     \multicolumn{2}{c}{450.5}&
			     \multicolumn{2}{c}{555.0}&
			     \multicolumn{2}{c}{668.4}\\
\hline
  \hline
\multicolumn{13}{c}{total eclipse}\\
\hline
$N$           & 4 && 3 && 3 && 3 && 8 && 3 &\\   
$r = 5\arcsec$	\hspace*{-1mm}&\hspace*{-1mm}$2.0 \pm 0.2$&[1.7, 2.3]\hspace*{-1mm}&\hspace*{-1mm}$2.3 \pm 0.6$&[2.0, 3.0]\hspace*{-1mm}&\hspace*{-1mm}$3.4 \pm 0.9$&[2.8, 4.4]\hspace*{-1mm}&\hspace*{-1mm} 
			  	$2.9 \pm 0.4$&[2.4, 3.2]\hspace*{-1mm}&\hspace*{-1mm}$4.7 \pm 1.5$&[1.7, 5.9]\hspace*{-1mm}&\hspace*{-1mm}$5.3 \pm 1.5$&[3.6, 6.4]\\
$r =  10\arcsec$	\hspace*{-1mm}&\hspace*{-1mm}$1.5 \pm 0.2$&[1.2, 1.7]\hspace*{-1mm}&\hspace*{-1mm}$1.9 \pm 0.5$&[1.6, 2.4]\hspace*{-1mm}&\hspace*{-1mm}$2.9 \pm 0.7$&[2.4, 3.7]\hspace*{-1mm}&\hspace*{-1mm} 
				$2.3 \pm 0.3$&[2.0, 2.5]\hspace*{-1mm}&\hspace*{-1mm}$4.2 \pm 1.5$&[1.1, 5.3]\hspace*{-1mm}&\hspace*{-1mm}$4.7 \pm 1.5$&[3.0, 5.7]\\
$r =  20\arcsec$\hspace*{-1mm}&\hspace*{-1mm}$0.8 \pm 0.6$&[0.0, 1.3]\hspace*{-1mm}&\hspace*{-1mm}$1.5 \pm 0.4$&[1.3, 2.0]\hspace*{-1mm}&\hspace*{-1mm}$2.6 \pm 0.6$&[2.0, 3.2]\hspace*{-1mm}&\hspace*{-1mm} 
				$1.9 \pm 0.2$&[1.7, 2.1]\hspace*{-1mm}&\hspace*{-1mm}$3.8 \pm 1.4$&[0.8, 4.8]\hspace*{-1mm}&\hspace*{-1mm}$4.4 \pm 1.4$&[2.8, 5.2]\\      
\hline
\multicolumn{13}{c}{partial eclipse}\\
\hline
$N$             &$\ldots$&&$\ldots$&& 9           &          &$\ldots$&&$\ldots$&&$\ldots$&\\   
$r =   5\arcsec$&$\ldots$&&$\ldots$&&$2.0 \pm 0.8$&[1.1, 3.2]&$\ldots$&&$\ldots$&&$\ldots$&\\
$r =  10\arcsec$&$\ldots$&&$\ldots$&&$1.6 \pm 0.7$&[0.9, 2.7]&$\ldots$&&$\ldots$&&$\ldots$&\\
$r =  20\arcsec$&$\ldots$&&$\ldots$&&$1.4 \pm 0.6$&[0.7, 2.2]&$\ldots$&&$\ldots$&&$\ldots$&\\
  \hline
\multicolumn{13}{c}{Mercury transit at solar disc-centre}\\
\hline
$N$           & 8  &&  5 &&  5 &&  5 &&  5 && 4  &\\   
$r =   5\arcsec$\hspace*{-1mm}&\hspace*{-1mm}$5.9 \pm 0.9$&[4.9, 6.7]\hspace*{-1mm}&\hspace*{-1mm}$ 6.2 \pm 0.3$&[ 5.7,  6.5]\hspace*{-1mm}&\hspace*{-1mm}$ 8.1 \pm 0.9$&[ 7.1,  9.2]\hspace*{-1mm}&\hspace*{-1mm} 
				$7.0 \pm 0.7$&[6.2, 8.1]\hspace*{-1mm}&\hspace*{-1mm}$10.3 \pm 0.2$&[10.2, 10.6]\hspace*{-1mm}&\hspace*{-1mm}$11.3 \pm 0.5$&[10.7, 12.0]\hspace*{-1mm}\\
  \hline
\multicolumn{13}{c}{Mercury transit at solar limb}\\
\hline
$N$         &  $\ldots$ &&   $\ldots$&&   $\ldots$&&   $\ldots$&& 39  && $\ldots$  &\\   
$r =   5\arcsec$&&&&&&&&&\hspace*{-1mm}$8.0 \pm 2.2$&[6.3, 13.8]\hspace*{-1mm}&&\\
  \hline
\multicolumn{13}{c}{solar limb}\\
\hline
$N$             &$\ldots$&&$\ldots$&&$\ldots$&& 66  &&  18 &&  20 &\\   
$r =   5\arcsec$&$\ldots$&&$\ldots$&&$\ldots$&\hspace*{-1mm}&\hspace*{-1mm} $5.3 \pm 0.7$&[4.3, 6.8]\hspace*{-1mm}&\hspace*{-1mm}$6.9 \pm 1.1$&[5.7, 10.5]\hspace*{-1mm}&\hspace*{-1mm}$7.9 \pm 1.3$&[5.5, 9.9]\hspace*{-1mm}\\
$r =  10\arcsec$&$\ldots$&&$\ldots$&&$\ldots$&\hspace*{-1mm}&\hspace*{-1mm} $4.7 \pm 0.6$&[3.7, 6.1]\hspace*{-1mm}&\hspace*{-1mm}$6.4 \pm 1.2$&[5.1, 9.9] \hspace*{-1mm}&\hspace*{-1mm}$7.2 \pm 1.2$&[5.0, 9.1]\hspace*{-1mm}\\
$r =  20\arcsec$&$\ldots$&&$\ldots$&&$\ldots$&\hspace*{-1mm}&\hspace*{-1mm} $4.0 \pm 0.7$&[2.9, 5.7]\hspace*{-1mm}&\hspace*{-1mm}$6.0 \pm 1.3$&[3.6, 9.0] \hspace*{-1mm}&\hspace*{-1mm}$6.7 \pm 1.1$&[5.0, 8.3]\hspace*{-1mm}\\    
  \hline
\hline 
  \end{tabular}
\end{table*}

The position of the FOV and individual subregions on the solar disc is here reduced 
to the heliocentric position \mbox{$\mu = \cos \theta$}.  
No distinction is made between the different parts of the limb (N/W/S/E) as there 
is no noticeable difference in the limb profiles for the quiet Sun 
\citep[see, e.g.][]{2002A&A...382..312L}. 
The pixel scale of the detector is 0\,\farcs05448 and thus smaller than the 
spatial resolution (and even the diffraction limit) of the optical system 
\citep[0\,\farcs2 - 0\,\farcs3, see, e.g.,][and Sect.~\ref{sec:intro}]
{2007SoPh..243....3K}.
As the same is still true for images with $2 \times 2$ pixel-binning, no difference in 
terms of contrast etc. is to be expected. 
Both binned and unbinned images are thus used next to each other. 

The data reduction (incl. dark subtraction and flat-fielding) was carried out with
Solar Soft \citep{2000eaa..bookE3390F}.
Dark frames and flat-fields were kindly provided by T.~Berger. In a few cases 
remaining hot pixels, which often occur in compact groups of adjacent pixels, had to 
be removed. 

\begin{figure*}[t]
   \sidecaption
   \centering
   \includegraphics[width=12cm]{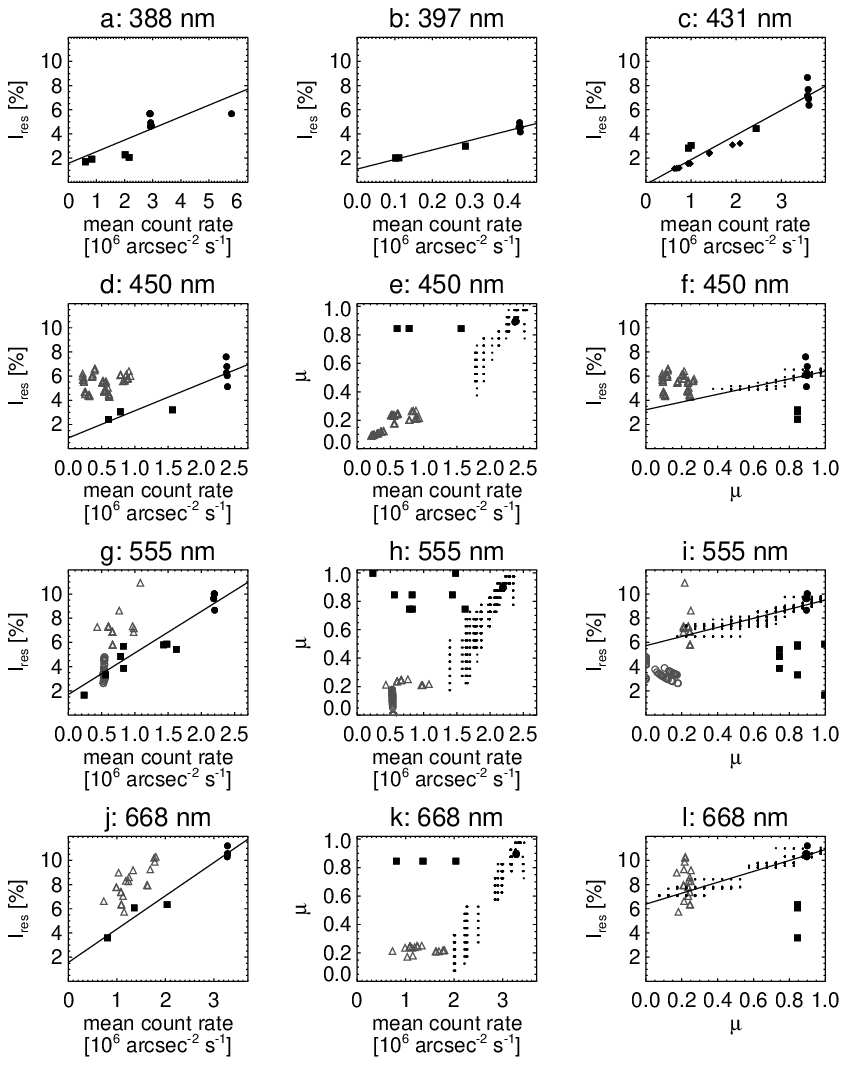}
   \caption{Residual intensity as function of mean count rate at a distance of 
   5\,\arcsec\ from the (eclipse) terminator, which corresponds to the centre of 
   the Mercury disc. 
   The panels in the upper row (\textbf{a}-\textbf{c} and the leftmost column 
   (\textbf{d}, \textbf{g}, \textbf{j}) show the residual intensity 
   $I_\mathrm{res, \lambda}$ as function of the mean count rate $\mathcal{C}$ for 
   the individual images of the Mercury transit (circles), the total eclipse 
   (squares), the partial eclipse (diamonds, 431\,nm only), 
   and the solar limb (triangles) at all six BFI bands. 
   Symbols are filled for $\mu > 0.5$ and unfilled and dark-grey otherwise.
   The adopted $\mu$-value for a data point is the average over the whole 
   individual image. 
   A image of the solar limb thus can have $\mu < 0.0$.  
   A relation between residual intensity and mean count rate is derived by 
   linear regression (solid lines), where the solar limb data is neglected. 
   It is justified by the uncertainties in choosing the right reference 
   intensity (see text for details).  
   More information is shown for the three wavelengths 
   450\,nm, 555\,nm, and 668\,nm.
   First the heliocentric position $\mu$ is plotted as function of 
   $\mathcal{C}$ in the middle column (\textbf{e}, \textbf{h}, \textbf{k} but 
   not \textbf{b}) for the eclipse and Mercury transit data and for images of 
   ``regular'' quiet Sun (dots). 
   The relation between $I_\mathrm{res, \lambda}$ and $\mathcal{C}$ is then 
   used to estimate $I_\mathrm{res, \lambda}$ from the mean count rate of 
   the regular (unocculted) images. The results are presented in the right 
   column (\textbf{f}, \textbf{i}, \textbf{l}). As abscissa $\mu$ is chosen. 
   Finally a linear regression is performed for the data points of regular 
   granulation (solid lines).}
   \label{fig:sotcnt}
\end{figure*}

\paragraph{Residual intensities.}
\label{sec:obsprofiles}
The intensity profiles and residual intensities in the dark areas are determined 
for the filtergrams from the Mercury transit, the two eclipses, and from limb 
observations. 
The residual intensity includes contributions from the PSF wings due to the optical 
system but also contributions from scattered light. 
All results are summarised in Table~\ref{tab:iresobs} for different distances from 
the terminator (for the eclipses) and the solar limb, respectively. 
For the Mercury transit, the residual intensity at its disc-centre is given 
($r \approx 5\arcsec$).  
The table contains the values for $I_\mathrm{res}$, which are the averages over all 
profiles, the corresponding standard variation, and (in brackets) the extrema. 
$N$ is the number of considered images.
The intensity profiles are averaged parallel to the terminator/limb, 
resulting in a radial average for the Mercury transit.  

Figure~\ref{fig:sotstray} illustrates the determination of the residual light level 
for the case of the green continuum. 
For the partial eclipse, however, the G-band is chosen because no other observations 
are available.
The right column displays the intensity profiles derived from the images on the left
(black solid lines).  
In addition also the corresponding profiles from all other images for the same 
wavelength are plotted as grey lines in each panel. 
All profiles are averaged over pixels with the same distance to the lunar/solar/Mercury 
limb, respectively. 
There is a general trend of a higher residual intensity with increasing wavelength, 
although the mean G-band value is slightly larger than for the neighbouring bands 
(see Table~\ref{tab:iresobs}). 
The variation of $I_\mathrm{res, \lambda}$ for images of the same wavelength is 
significant, clearly demonstrating that it is impossible to describe the  
stray-light contribution with a single number (for each wavelength). 
The situation is much better for the Mercury transit at solar disc-centre as 
the individual images were taken under very similar conditions, resulting in less 
variation of $I_\mathrm{res}$.
In the following the individual sets of images are characterised. 

\paragraph{Total eclipse.}
The mean residual intensities in the images for the total solar eclipse at a 
distance of 5\,\arcsec from the terminator amount to $(2.9 \pm 0.4)$\,\% of the 
mean intensity of the unocculted solar disc for the blue continuum but to 
$(5.3 \pm 1.5)$\,\% for the red continuum, respectively. 
At a distance of 20\,\arcsec the values reduce to $(1.9 \pm 0.2)$\,\% and 
$(4.4 \pm 1.4)$\,\%. 
Values between 0.8\,\% and 5.2\,\% are found for individual (averaged) profiles.

\paragraph{Partial eclipse.}
For the partial eclipse only G-band images are available. The intensity profiles 
essentially form two groups, one with a residual intensity at a distance 
of 20\,\arcsec of 0.7\,\% to 1.0\,\% of 
the solar disc intensity and the other with 1.8 to 2.2\,\%. 
The higher values result from images where the dark occulted area is less than 42\,\% 
of the image, resulting in a higher mean count rate (see below). 
The low values come from images where the lunar disc covers more than half of the 
image and the mean count rate is low. 

\paragraph{Mercury transit at solar disc-centre.}
Most images of the Mercury transit on the solar disc are taken at a heliocentric 
position of the Mercury disc of $\mu \approx 0.9$. 
The residual intensity at Mercury disc-centre is determined from the radially 
averaged profile across the Mercury limbwith respect to the mean of surrounding 
solar disc area of comparable $\mu$. 
Close to solar disc-centre ($\mu \approx 0.9$) the following values are derived 
for the three continuum bands: 
$(7.0\,\pm\,0.7)$\,\% (blue), 
$(10.3\,\pm\,0.2)$\,\% (green), and
$(11.3\,\pm\,0.5)$\,\% (red). 
For the other three broadband channels the following values are found: 
CN (388\,nm): $(5.9\,\pm\,0.9)$\,\%, 
Ca (397\,nm): $(6.2\,\pm\,0.3)$\,\%, and
Gb (431\,nm): $(8.1\,\pm\,0.9)$\,\%. 
The Mercury transit at solar disc-centre obviously gives residual intensities that 
are much higher than estimated from the eclipses and from the solar limb observations. 
This can already be seen from the intensity profiles in \mbox{Fig.~\ref{fig:sotstray}}, 
which decrease to much lower residual intensity values off the solar limb than compared 
to the centre of the Mercury disc. 
This is partly due the influence of the wings of the telescope's PSF, which contribute 
more to the residual intensity for the Mercury disc than for an eclipse 
(see Sect.~\ref{sec:psfexp}).  

\paragraph{Mercury transit at the solar limb.}
In addition there are green continuum images at the solar limb covering the beginning 
of the Mercury transit. 
The value of $I_\mathrm{res, \lambda}$ referres to the mean intensity of the 
solar disc close to the Mercury disc in order to ensure a comparable $\mu$. 
Next to the radial profiles also profiles along the x- and y-axis through the 
Mercury disc-centre are shown in the left part of \mbox{Fig.~\ref{fig:sotstray}j}. 
All include also the intensity decrease over the solar limb. 
At Mercury disc-centre the radial averaged profiles exhibit residual intensities of 
$(8.0\,\pm\,2.2)$\,\% with extreme values of 6.3\,\% and 13.8\,\%. 
Only the inner parts, which are of interest here, of the radial profiles are reliable 
as further away from the disc-centre one averages the non-radial solar disc. 
The averages along the $x$- and $y$-axes show very similar residual intensities 
(see \mbox{Fig.~\ref{fig:sotstray}j)}. 
Nevertheless the derived values for $I_\mathrm{res, \lambda}$ are smaller than for 
the Mercury transit at disc-centre because now a larger area of the image is dark, 
mapping the space beyond the solar limb. 
The large variation in $I_\mathrm{res, \lambda}$ can be explained by (i)~the 
varying fraction of the dark area and thus the overall light level in the telescope
and (ii)~the more problematic determination of a reference intensity as the Mercury 
disc at the solar limb spans a larger range in $\mu$.  

\paragraph{Solar limb.}
The pure limb observations (i.e. without Mercury, see Fig.~\ref{fig:sotstray}a-b) 
imply values for the residual intensity from $(5.3 \pm 0.7)$\,\% for blue continuum 
to $(7.9 \pm 1.3)$\,\% for the red continuum at a distance of 5\,\arcsec from 
the solar limb. For 20\,\arcsec the values decrease to $(4.0 \pm 0.7)$\,\% 
and $(6.7 \pm 1.1)$\,\%, respectively.  
The residual intensity refers to the mean intensity of the solar disc close to the 
limb (average running from $-1$\,\arcsec to $-2$\,\arcsec). The reference, however, 
is problematic as the limb images often extend to $\mu \sim 0.4$ and in some cases 
even to $\mu > 0.5$.

\subsection{Dependence on overall light level} 

The large variation in the eclipse intensity profiles for the same wavelength is 
caused by the different illumination of the detector during the exposure.
This ``overall light level'' is here quantified as the ``mean count rate'' 
$\mathcal{C}$. 
It is defined as the total count number of all pixels, i.e. the full FOV, 
of an image divided by the FOV area (in arcsec$^2$) and the exposure time. 
For a given wavelength, an unocculted image taken at solar disc-centre has a higher 
value of $\mathcal{C}$ than an image taken closer to the solar limb. 
$\mathcal{C}$ is reciprocally proportional to the fraction of the occulted and 
thus dark image area. 
For the images considered in this section, $\mathcal{C}$ is highest for the Mercury 
transit at disc-centre.
In all other cases a significant part of the detector is dark as it images the 
lunar disc or the background beyond the solar limb. 
The same is true for the Mercury transit images at the solar limb.  

The residual intensities are measured for all images at a distance of 5\arcsec\   
from the terminator/solar limb. 
The results for the different kinds of observations are represented 
by different symbols in Fig.~\ref{fig:sotcnt}. 
For all six wavelength channels $I_\mathrm{res, \lambda}$ increases linearly with 
$\mathcal{C}$. 
The eclipse and Mercury transit data was used to derive the regression lines 
shown in Fig.~\ref{fig:sotcnt}. 
The solid lines show the result of a linear regression, which reproduce the relation 
quite well. 
The solar limb profiles are excluded here because of the 
uncertain reference intensity. 
The solar limb data is in line with the eclipse and Mercury transit when 
plotting the corresponding count rate instead of $I_\mathrm{res}$. 
Obviously a higher reference intensity is necessary to make the limb data comply 
with the regression lines for $I_\mathrm{res}$. 
The dependence on $\mathcal{C}$ indicates that the residual intensity contains 
contributions from anisotropic stray-light, which, e.g. may originate from internal 
reflections in the instrument and thus clearly depend on the overall light level. 

The heliocentric positions $\mu$ are plotted as function of the mean count rate 
for the blue, green, and red continuum in the panels e, h, and k of 
Fig.~\ref{fig:sotcnt}, respectively. 
The panels f, i, and l show the residual intensity as function of $\mu$. 
The Mercury transit images for the green continuum have residual intensities of 
2.6\,\% to 4.8\,\% close to the solar limb \mbox{(position of the Mercury disc at 
$\mu < 0.2$)}, whereas $\sim 10.0$\,\% are derived close to disc-centre. 
It clearly demonstrates the increase of the relative residual intensity with $\mu$, 
related to the centre-to-limb variation of the mean intensity of solar granulation 
and the corresponding change of the mean count rate. 

The residual intensities in images of ``regular'' unocculted granulation are now 
estimated from the corresponding values of $\mathcal{C}$ and the (wavelength-dependent)  
relation with $I_\mathrm{res, \lambda}$, which was found from the Mercury transit and 
the eclipse data. 
The results for the blue, green, and red continuum are represented as dots in 
Fig.~\ref{fig:sotcnt}. 
Finally a linear regression of the form $I_\mathrm{res, \lambda} = m \cdot \mu + c$ 
is performed for each wavelength (solid lines in Fig.~\ref{fig:sotcnt}).
The derived parameters $c$ (constant) and $m$ (slope) are listed together with the 
maximum value of $I_\mathrm{res, \lambda}$ at $\mu = 1.0$ for all three considered 
continuum wavelengths in Table~\ref{tab:iresest}. 
At solar disc-centre the following residual intensities $I_\mathrm{res, \lambda}$ 
can be expected in images of ``regular'' granulation: 
6.4\,\%, 9.5\,\%, and 10.9\,\%, for blue, green, and red continuum, respectively. 
The estimate is based on residual intensities at 5\,\arcsec distance from the 
terminator of the Mercury transit and the eclipses and should thus be interpreted 
as upper limit only. 

\begin{table}[pb]
  \caption[h]{Estimate of the residual intensity $I_\mathrm{res, \lambda}$ 
  for images of ``regular'' granulation in percent points. See text for details.}
  \label{tab:iresest}
  \centering
  \begin{tabular}[b]{cccc}
  \hline
  \hline
  wavelength&$m$ [\%]&$c$ [\%]&$(I_\mathrm{res, \lambda})_\mathrm{max}$ [\%]\\
  \hline
  450.45&3.18& 3.22 & 6.40\\
  555.00&3.77& 5.73 & 9.50\\
  668.40&4.51& 6.39 &10.90\\
  \hline
  \hline
  \end{tabular}
\end{table}
\begin{figure}[t]
  \centering
  \includegraphics{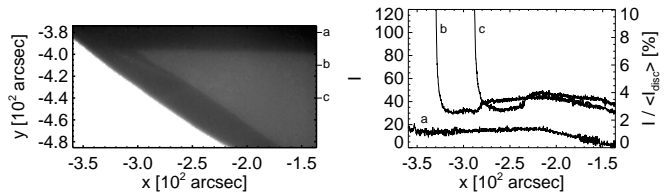}
   \caption{Stray-light in a SOT  green continuum filtergram of the total 
   eclipse from March 19th, 2007 ($0.78 < \mu < 0.92$): 
   Image (left) and profiles along the x-axis (right). 
   In order to make the stray-light visible, the data range is limited to a value 
   of 120, which is about 10\,\% of the mean value of the bright solar disc (lower 
   left corner of the image).
   The vertical positions of the profiles are indicated by tick marks on the right 
   side of the image. 
   Profile a lies in a "shadow", whereas profile b crosses a "streak" at a position 
   of $x = -300$\,\arcsec.}
   \label{fig:sotstrayimg}
\end{figure}

\subsection{Anisotropy} 
A direct proof of stray-light is presented in Fig.~\ref{fig:sotstrayimg}. 
It shows a frame taken during the total eclipse but with the displayed intensity 
range limited to only 10\,\% of the mean value of the bright solar disc.  
There are reflections (e.g. from baffles) and a shadow at the upper edge of the 
detector. 
More ``artefacts'' that are not removed during the data reduction process can be 
found. 
For instance, there is a streak-like feature close to the terminator. 
Obviously the stray-light is highly anisotropic. 
All those details are relatively subtle and might be negligible for most 
applications for which exact photometry is not needed. 
Nevertheless the residual intensity can reach up to 4\,\% of the solar disc 
intensity in this example. 
Similar effects are expected for the other channels, too. 
For instance, the red continuum channel of the BFI shows a prominent spot. 
Variations in the stray-light contribution due to asymmetric scattering inside the 
optical system have been reported for other telescopes before 
\citep[e.g.][]{1983SoPh...87..187M}.

\section{Theoretical PSFs for SOT}
\label{sec:psfmodel}

\subsection{Ideal diffraction-limited PSFs}
\label{sec:psfideal}
The diffraction-limited PSF of a circular homogenous illuminated aperture of 
diameter $d$ is described by the Airy function. 
For a wavelength $\lambda$ the FWHM of the central Airy disc is equal to the 
resolution of the telescope\footnote{It should not be confused with the Rayleigh 
criterion, which is defined with an additional factor of 1.22 for the position of 
the first minimum surrounding the central Airy disc (approximately the first minimum 
of the Bessel function of the first kind).} 
\begin{equation}
\label{eq:difflimit}
\sin \alpha_\mathrm{res} = \frac{\lambda}{d}\enspace. 
\end{equation}

For a more complicated aperture geometry the PSF can be calculated as the square 
of the wave number Fourier transform of the pupil shape. 
It has to be normalised so that the sum is equal to one. 
For SOT the 50\,cm diameter aperture, the 17.2\,cm diameter central obscuration, and 
the three spiders ($120^\circ$ separation, 4\,cm wide) are taken into account. 
Each PSF is defined as an array with 5120$^2$ pixels and a spatial increment of half 
a SOT detector pixel. 
The extent of the PSFs is thus larger than the minor axis of the FOV of BFI images 
and includes the far wings. 
The central lobes have the following FWHMs (cf. Eq.~(\ref{eq:difflimit})): 
0\,\farcs18 at 450.45\,nm (blue), 
0\,\farcs22 at 555.05\,nm (green), and 
0\,\farcs27 at 668.40\,nm (red), respectively. 
\citet{suematsu07privcomm} kindly provided an ideal G-band PSF for the main 
mirror of SOT, which has a FWHM of 0\,\farcs18. 
As ideal PSFs can be scaled reciprocally with wavelength, the G-band PSF 
can be compared to the PSFs calculated here. 
They agree within the limits of numerical uncertainties. 

\begin{figure}[t]
  \includegraphics{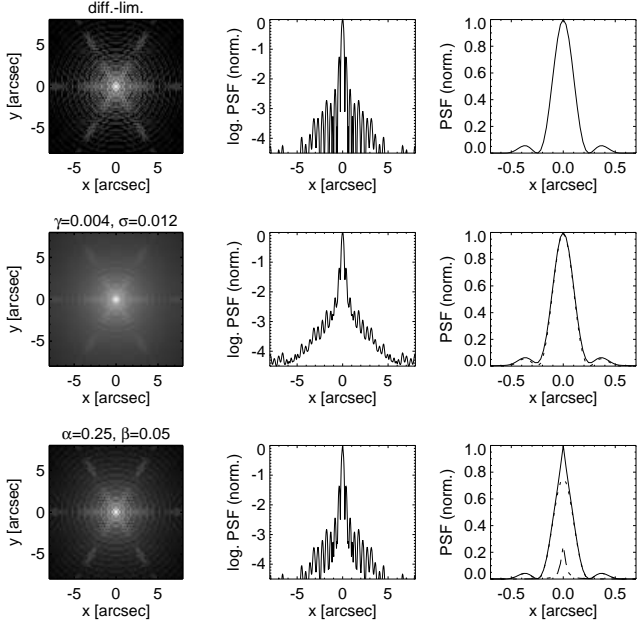}  
  \caption{Theoretical point spread functions for a wavelength of 555\,nm:
	  Original diffraction-limited PSF (top row), after 
	  convolution with a Voigt function with $\gamma =0\,\farcs004$ and 
	  $\sigma =0\,\farcs012$	(middle), and after linear combination with a 
	  Lorentzian with the parameters $\alpha = 0.25$ and $\beta = 0.05$ (bottom).   
	  The left columns shows the inner part of the PSFs on a logarithmic grey-scale, 
	  which covers six orders of magnitude, whereas the corresponding profiles along 
	  the x-axis are plotted in the middle column. The rightmost panels display the
	  the PSF core on a linear scale (solid) together with the original PSF (dotted). 
	  The dashed line in the lower right panel is the used Lorentzian profile.}
   \label{fig:sotpsf}
\end{figure}

\subsection{Non-ideal PSFs}
\label{sec:psfnonid}

Here Voigt functions are used as approximation for the non-ideal PSF contributions 
(see Sect.~\ref{sec:intro}). 
The final PSFs are then derived by a  
convolution of the ideal PSFs $\mathcal{P}_\mathrm{dl}$ and a Voigt 
function $\mathcal{V}_\gamma$
\begin{equation} 
	\mathcal{V}_{\gamma,\sigma}\,(r) 
	= 
	(2\pi)^{-3/2}\,\frac{\gamma}{\sigma} \,
	\int_{-\infty}^{+\infty} \frac{e^{-r'^2/(2 \sigma^2)}}{(\frac{\gamma}{2})^2 
	+ (r - r')^2} \, dr'\,,
\end{equation}
which is the convolution of a Gaussian 
\begin{equation} 
	\mathcal{G}_\sigma (r) = 
	\frac{e^{- r^2 / (2\,\sigma^2)}}{\sqrt{2\,\pi}\,\sigma}
\end{equation}
and a Lorentz function
\begin{equation}
	\mathcal{L}_\gamma\,(r) = 	
	\frac{\frac{\gamma}{2}}{\pi\, ( r^2 + \frac{\gamma^2}{4})}. 
\end{equation}
Here $r = \sqrt{x^2 + y^2}$  is the spatial ordinate and  
$\gamma$ is the FWHM of the Lorentzian. 
The FWHM of the Gaussian is defined as $2\,\sigma\,(2\,\ln{2})^{1/2}$. 
With 
$u = r /(\sqrt{2} \sigma)$ and the damping parameter 
\begin{equation}
	\label{eq:voigtdamp}
	a = \frac{\gamma}{2}\,\frac{1}{\sqrt{2}\,\sigma}
\end{equation}		 
the Voigt function can also be written in the form 
\begin{equation} 
	\mathcal{V} (a, u) 
	=
	\frac{1}{\sqrt{2\,\pi}\,\sigma}\,\frac{a}{\pi}\,
	\int_{-\infty}^{+\infty} \frac{e^{-{u'}^2}}{a^2 + (u - u')^2} \, du'
	= 
	\frac{H (a, u)}{\sqrt{2\,\pi}\,\sigma}.
\end{equation}
The far wings, i.e. the PSF at large angles, are normally dominated by 
the Voigt function. 
The function is numerically reduced to a delta peak if the FWHM of the 
Voigt function is significantly smaller than the numerical axis increment. 
In that case the convoluted PSF is identical to the original diffraction-limited 
one.
Here PSFs with an additional Voigt function $\mathcal{V}_{\gamma,\sigma}$ are 
computed for a parameter grid of $\gamma$ and~$\sigma$. 
The grid includes PSFs for which $\mathcal{P}_\mathrm{dl}$ is convolved 
with a pure Lorentzian $\mathcal{L}_\gamma$ for different values of~$\gamma$ or 
with a pure Gaussian $\mathcal{G}_\sigma$ for different values of~$\sigma$.

In some studies a linear combination of the ideal PSF and a Lorentzian of the 
type 
\begin{equation}
\label{eq:psflor}
\mathcal{P}_\mathrm{L}\,(x, y) \, = \, 
(1 - \alpha) \, \mathcal{P}_\mathrm{dl}\,(x, y) 
+ \alpha \,  \mathcal{L}_\beta\,( r(x, y) )
\end{equation}
is used instead of a convolution. 
The linear approach is considered here only for comparison as it is strictly speaking 
physically not correct. 
Instead the convolution method is preferrable (see Sect.~\ref{sec:discus}). 
To avoid confusion with the convolution approach the FHWM of the Lorentzian is 
here named $\beta$ instead of $\gamma$. 
A parameter grid for $[\alpha,\beta]$ is computed. 

\section{Fitting observations with synthetic profiles}
\label{sec:psfexp}

\begin{figure}[t]
	\centering
	\includegraphics{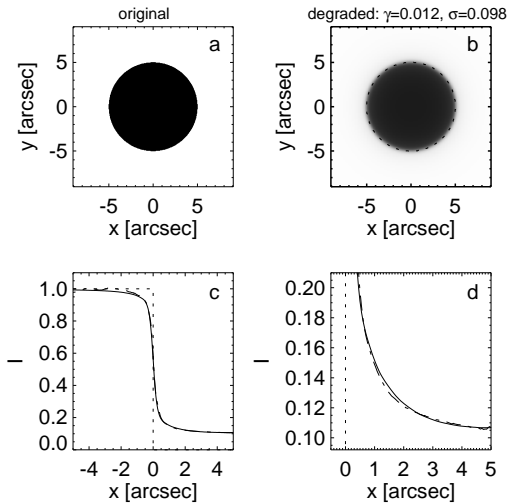}  
  	\caption{Artificial observation of a Mercury transit  
	at a wavelength of 555\,nm:   
	\textbf{a)} innermost region of the original intensity mask, 
	\textbf{b)} degraded image after convolution with a non-ideal PSF that was calculated by 
	a diffraction-limited PSF and a Voigt profile with \mbox{$\gamma = 0\,\farcs012$}
	and	\mbox{$\sigma = 0\,\farcs098$}, 
	\textbf{c)} radially averaged intensity profiles for the original image (dotted) and
	the degraded image (solid), and 
	\textbf{d)} close-up of panel~c.
	The dot-dashed line is an empirical intensity profile derived from a SOT image.}
	\label{fig:artobs}
\end{figure}

Here the different theoretical PSFs for SOT (see Sect.~\ref{sec:psfmodel}) are 
applied to artificial idealised two-dimensional images of a Mercury transit and 
an eclipse. 
The image dimensions conform to half of the BFI detector, i.e. $2048 \times 2048$
pixels with a scale of $0\,\farcs05448\,$px\,arcsec$^{-1}$, resulting in a field 
of 111\,\farcs6$\,\times\,$111\,\farcs6.
The resulting synthetic degraded intensity profiles are then compared to empirical 
profiles derived from SOT images (see Fig.~\ref{fig:artobs} for an example).
The optimum PSF is the one that produces the smallest difference between observed 
and synthetic intensity profile. 

\begin{figure*}[t]
	\centering
	\includegraphics{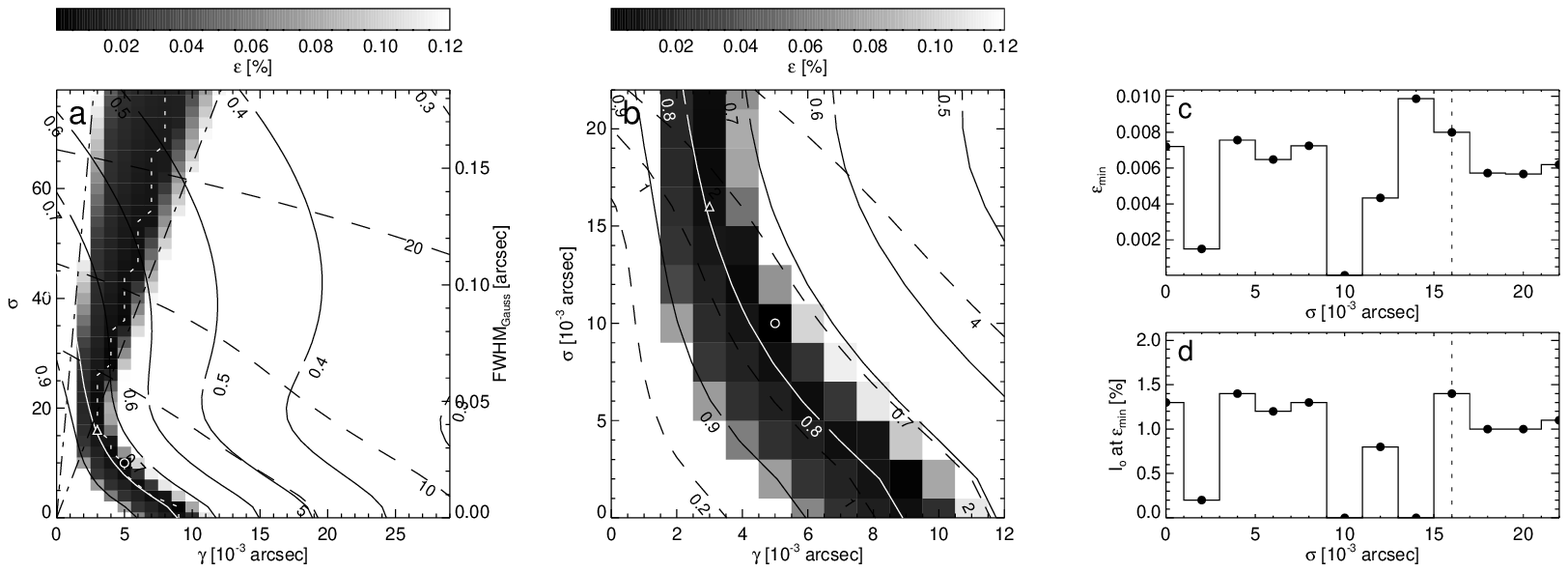}  
  	\caption{Goodness $\varepsilon$ of the fits for the reference case of 
	the Mercury transit at $\lambda = 555$\,nm. 
	It represents the difference between a prescribed synthetic intensity 
	profile for a parameter pair of $\gamma = 0\,\farcs005$ and 
	$\sigma = 0\,\farcs01$ and all other profiles. 
	They are derived with non-ideal PSFs calculated by convolution with a Voigt 
	function. 
	\textbf{a)} The goodness $\varepsilon$ as defined in 
	Eq.~(\ref{eq:psffitgood}) as function of $\gamma$ and $\sigma$, 
	\textbf{b)} close-up of the lower left corner, 
	\textbf{c)} $\varepsilon$ along the ``minimum error trench'' (MET, marked 
	with white dotted line in panel~a), and
	\textbf{d)} the applied intensity offset $I_\mathrm{o}$ along the MET.  
	Overplotted in panel a and b are contours of constant Strehl ratio 
	(solid, black and white) and of constant broadening (dashed) of the combined PSF.
	The broadening is given in percent points increase of the FWHM of the PSF with 
	respect to the diffraction-limited PSF.
	The circles in panel a and b mark the parameter pair with the smallest $\varepsilon$
	below the ``knee'' of the MET, while the ``knee'' is indicated by a triangle. 
	The dot-dashed lines in panel~a are for constant Voigt damping parameter $a$, 
	whereas the dotted vertical lines in panels c and d show the rough location 
	of the MET ``knee''.}
	\label{fig:psffitref}
\end{figure*}

\subsection{Ideal diffraction-limited PSFs}
\label{sec:artlimbideal}

\begin{table}[b]
\caption[h]{Residual intensities $I_\mathrm{res}$ in the artificial 
diffraction-limited$^{\mathrm{a}}$ and the SOT observations$^{\mathrm{b}}$.}
\label{tab:dfpsfires}
\begin{center}
\begin{tabular}{cl|ccc}
\hline
$I_\mathrm{res}$ [\,\%]&&450\,nm&555\,nm&668\,nm\\
\hline
\multicolumn{5}{c}{Mercury disc-centre ($r = 5$\,\arcsec)}\\
\hline
$r = 5$\,\arcsec&
observed   & 7.0 $\pm$ 0.7& 10.3 $\pm$ 0.2& 11.3 $\pm$ 0.5\\
&theoret.  & 1.8         &  2.2                 &  2.6 \\
&difference&5.2 $\pm$ 0.7&  8.1 $\pm$ 0.2&  8.7 $\pm$ 0.5\\
\hline
\multicolumn{5}{c}{total eclipse, different distances $r$ from lunar limb}\\
\hline
$r = 5$\,\arcsec&
observed   & 2.9 $\pm$ 0.4& 4.7 $\pm$ 1.5& 5.3 $\pm$ 1.5\\
&theoret.  & 0.6          & 0.7          & 0.8\\ 
&difference& 2.3 $\pm$ 0.4& 4.0 $\pm$ 1.5& 4.5 $\pm$ 1.5\\
\hline
$r = 10$\,\arcsec&
observed   & 2.3 $\pm$ 0.3& 4.2 $\pm$ 1.5& 4.7 $\pm$ 1.5\\
&theoret.  & 0.3          & 0.3          & 0.4\\
&difference& 2.0 $\pm$ 0.3& 3.9 $\pm$ 1.5& 4.3 $\pm$ 1.5\\
\hline$r = 20$\,\arcsec&
observed   & 1.9 $\pm$ 0.2& 3.8 $\pm$ 1.4& 4.4 $\pm$ 1.4\\  
&theoret.  & 0.2          & 0.2          & 0.3          \\
&difference& 1.7 $\pm$ 0.2& 3.6 $\pm$ 1.4& 4.1 $\pm$ 1.4\\
\hline
\end{tabular}
\end{center}
\begin{list}{}{}
\item[$^{\mathrm{a}}$] 
All theoretical values are corrected for the influence of the far PSF wings at different distances from the terminator (see Sect.~\ref{sec:dfartmer}). 
\item[$^{\mathrm{b}}$] 
The difference between both is due to 
contributions not included in the ideal PSFs, e.g., stray-light inside the 
telescope.
\end{list}
\end{table}

\subsubsection{Artificial Mercury transit}
\label{sec:dfartmer}
The ideal theoretical PSFs for each wavelength are applied to a simple binary mask of 
a bright solar background with a black Mercury disc (cf. Fig.~\ref{fig:artobs}).
Convolution with the PSF transforms the sharp edge of the Mercury limb in the initial 
mask into a smooth slope. 
The ideal SOT PSF for a wavelength of 450.1\,nm (blue continuum) produces a residual 
intensity $I_\mathrm{res}$ of 1.7\,\% (with respect to the bright solar background) 
at the centre of the Mercury disc in the degraded image. 
A simpler Airy PSF results in only 0.7\,\% residual intensity, implying a substantial 
contribution due to the spiders and the central obstruction of the telescope. 
Already replacing the SOT PSF with its radial mean, i.e., removing the non-radial 
variation due to the spiders, significantly decreases the residual intensity. 
Quantitative results obviously demand for a detailed model of the telescope 
properties.

Although the considered FOV is already relatively large, the residual intensity 
still grows asymptotically as a function of image (and PSF) size. 
The far wings of the PSF contribute only little to the intensity but nevertheless 
produce an asymptotic increase of $I_\mathrm{res}$ (at disc-centre) by a factor 
of $\sim 1.1$ with respect to a field size of 
111.6\,\arcsec$\,\times\,$111.6\,\arcsec.  
Approximately the same factor is found for $\lambda = 555.0$\,nm and 668.4\,nm.
The estimate is based on an analytical 2D Airy function.
Under the assumption that the detailed PSFs show the same qualitative behaviour, 
the factor 1.1 is now multiplied to the corresponding values of $I_\mathrm{res}$, 
resulting in  residual intensities of 
1.8\,\% (450.1\,nm), 2.2\,\% (555.0\,nm), and 2.6\,\% (668.4\,nm), respectively
(see Table~\ref{tab:dfpsfires}). 
The proportional increase with wavelength is related to the broadening 
of the PSF (see the FWHM).

\subsubsection{Artificial eclipse}
The experiment is repeated for artificial eclipse images with a straight 
terminator. 
Different field sizes and also positions of the limb within the images are 
considered. 
Again the residual intensity, which is derived with the ideal PSF, is 
corrected for the asymptotic behaviour for very large bright areas due to the far 
wings. 
At a distance of 5\,\arcsec, which corresponds to the Mercury disc radius, the 
following residual intensities (corrected with a factor of 1.1 for the far wings) 
are found to be 0.6\,\%, 0.7\,\%, and 0.8\,\% for blue, green, and red continuum, 
respectively (see Table~\ref{tab:dfpsfires}).
The values are much smaller than for the Mercury experiment owing to the different 
geometrical situations and the different proportions of bright to occulted area. 
In contrast to the Mercury transit experiment, where the geometry is 
radial-symmetric and the occulted area is relatively small, the eclipse has a 
clear anisotropic geometry with a significant occulted area. 
At a location beyond the eclipse terminator only radiation from one side is 
received, whereas the centre of the Mercury disc is irradiated by the bright 
(solar) disc from all directions.  

\subsubsection{Comparison with observations} 
Comparison of the observed and artificial intensity profiles show that when 
using an ideal PSF 
(i)~the average observed intensity profiles over the terminator have a less steep 
gradient  and (ii)~the residual intensities derived from the observations are 
generally larger (see Table~\ref{tab:dfpsfires}). 
This indicates that (i)~the true PSFs are broader than the theoretical 
diffraction-limited ones and (ii)~that there is an additional contribution due 
to stray-light.  
The difference between theoretical and observed profiles decreases with distance 
from the terminator and can thus not be described with just a constant intensity 
offset.  
Many profiles, averaged parallel to the terminator or just individual profiles, 
exhibit even a change in slope. 
The slope can get steeper but also reverse again in some cases (see 
Figs.~\ref{fig:sotstray} and \ref{fig:sotstrayimg}). 
The synthetic degraded intensity profiles would fit the terminator slope of the 
observed (averaged) profiles better if the PSFs were broader. 
A broader PSF corresponds to a smaller effective aperture (or a longer wavelength, 
see Eq.~(\ref{eq:difflimit})). 
However, even though a broader PSF would provide a better fit of the slope, the 
(dark) part of the profiles directly beyond the terminator is often not well 
reproduced. 
Many observed profiles do not show the theoretical asymptotic  decrease towards a 
residual intensity, indicating that a purely diffraction-limited PSF is too simple. 

\begin{table*}
  \caption[h]{Parameters$^{\mathrm{a}}$ of the convolved non-ideal PSFs that produce 
  the best fit of the intensity profiles across the Mercury disc and the total 
  eclipse.}
  \label{tab:psffit}
  \centering
  \begin{tabular}[b]{cccc|ccc}
  \hline
  \hline
  	$\lambda$&
  		450.45\,nm&555.00\,nm&668.40\,nm&
  		450.45\,nm&555.00\,nm&668.40\,nm\\
  	\hline\ \\[-2mm]
	&
	\multicolumn{3}{c}{Mercury transit}&
	\multicolumn{3}{c}{Eclipse\vspace*{1mm}}\\
  	\hline
	\multicolumn{7}{c}{Convolution with Voigt function: best fit\,$^{\mathrm{b}}$}\\
	\hline	
	$\gamma_\mathrm{opt}$      &$    4 \ (4 \pm 2)$&$   5 \ (6 \pm 2)$&$   6 \ (6 \pm 2)$ 
	              &$    5 \ (5 \pm 2)$&$   5 \ (4 \pm 2)$&$   4 \ (4 \pm 1)$\\
	$\sigma_\mathrm{opt}$      &$    8 \ (8 \pm 5)$&$   8 \ (8 \pm 4)$&$   8 \ (8 \pm 5)$
	              &$    7 \ (7 \pm 5)$&$   7 \ (7 \pm 3)$&$   7 \ (7 \pm 3)$\\	
	$I_\mathrm{o}$&$ 0.3  \ (0.5 \pm 0.4)$&$0.4 \ (0.3 \pm 0.3) $&$0.4 \ (0.4 \pm 0.3)$
	              &$ 0.1  \ (0.2 \pm 0.2)$&$2.5 \ (2.7 \pm 0.4) $&$3.0 \ (3.2 \pm 0.3)$\\
	$b$           &$  0.8 \ (0.8  \pm 0.2) $&$ 1.4 \ (1.4 \pm 0.1)  $&$ 1.3 \ (1.3  \pm 0.1)$
	              &$  0.9 \ (0.8  \pm 0.2) $&$ 0.9 \ (0.8 \pm 0.3)  $&$ 0.8 \ (0.7  \pm 0.2)$\\
	$S$           &$ 0.84 \ (0.84 \pm 0.03)$&$0.78 \ (0.78 \pm 0.01)$&$0.78 \ (0.78 \pm 0.01)$
	              &$ 0.83 \ (0.84 \pm 0.02)$&$0.84 \ (0.86 \pm 0.04)$&$0.85 \ (0.87 \pm 0.03)$\\
  	\hline
	\multicolumn{7}{c}{Linear regression of MET below the ``knee''\,$^{\mathrm{c}}$}\\
	\hline
	$\Delta\sigma/\Delta\gamma$&
		       $-2.672 \pm 0.218 $&$-2.228 \pm 0.164 $&$-2.277 \pm 0.148$
	          &$-2.587 \pm 0.309 $&$-2.528 \pm 0.237 $&$-2.619 \pm 0.097$\\ 
	$\sigma_C$&$  18.2 \pm 1.6   $&$  20.4 \pm 1.7   $&$  21.0 \pm 1.5$
	          &$  18.7 \pm 0.9   $&$  17.7 \pm 3.2   $&$  17.8 \pm 2.1$\\
	$b$       &$   0.8 \pm 0.2   $&$1.4    \pm 0.3   $&$1.2    \pm 0.2$
	          &$   0.9 \pm 0.2   $&$0.9    \pm 0.3   $&$0.8    \pm 0.2$\\
	$S$       &$  0.85 \pm 0.03  $&$0.78   \pm 0.03  $&$0.79   \pm 0.03$
	          &$  0.84 \pm 0.03  $&$0.86   \pm 0.04  $&$0.86   \pm 0.03$\\ 
    \hline
	\multicolumn{7}{c}{Convolution with Voigt function: at MET ``knee''\,$^{\mathrm{d}}$}\\
	\hline	
	$\gamma$&$         3 \pm 1   $&$   4 \pm   1 $&$   4 \pm 1   $&
	         $         3 \pm 1   $&$   3 \pm   1 $&$   3 \pm 1   $\\
	$\sigma$&$        13 \pm 2   $&$  12 \pm   1 $&$  12 \pm 1   $&
             $        14 \pm 2   $&$  12 \pm   2 $&$  13 \pm 5   $\\ 
	$I_\mathrm{o}$&$ 0.4 \pm 0.3 $&$ 0.4 \pm 0.6 $&$ 0.4 \pm 0.3 $&
	               $ 0.1 \pm 0.2 $&$ 2.5 \pm 0.4 $&$ 3.0 \pm 0.4 $\\ 
	$b$     &$       0.9 \pm 0.2 $&$ 1.5 \pm 0.2 $&$ 1.4 \pm 0.2 $&
	               $ 1.0 \pm 0.1 $&$ 1.1 \pm 0.5 $&$ 0.9 \pm 0.2$\\
	$S$     &$      0.84 \pm 0.02$&$0.78 \pm 0.01$&$0.78 \pm 0.02$&
	               $0.83 \pm 0.02$&$0.84 \pm 0.05$&$0.85 \pm 0.04$\\ 
	\hline
	\multicolumn{7}{c}{Convolution with Lorentzian ($\sigma = 0$)\,$^{\mathrm{d}}$}\\
	\hline
	$\gamma$      &$   7 \pm 1   $&$   9 \pm 1   $&$   9 \pm 1$&
                   $   8 \pm 1   $&$   7 \pm 1   $&$   7 \pm 1$\\
	$I_\mathrm{o}$&$ 0.8 \pm 0.7 $&$ 0.6 \pm 0.3 $&$ 0.9 \pm 0.5$&
	               $ 0.1 \pm 0.2 $&$ 2.7 \pm 0.4 $&$ 3.2 \pm 0.4$\\
	$b$           &$ 0.7 \pm 0.1 $&$ 1.3 \pm 0.1 $&$ 1.2 \pm 0.1$& 
	               $ 0.8 \pm 0.1 $&$ 0.8 \pm 0.3 $&$ 0.7 \pm 0.2$\\
	$S$           &$0.86 \pm 0.02$&$0.79 \pm 0.02$&$0.79 \pm 0.02$&
	               $0.84 \pm 0.02$&$0.86 \pm 0.04$&$0.87 \pm 0.03$\\
  	\hline
  	\hline
  \end{tabular}
  \begin{list}{}{}
  \item[$^{\mathrm{a}}$] 
  $\gamma$ and $\sigma$ in units of $10^{-3}$\,\arcsec, 
  intensity offset $I_\mathrm{o}$ in \%, 
  broadening factor $b$ of the FWHM with respect to the ideal PSF in \%, 
  and the Strehl ratio $S$.
  \item[$^{\mathrm{b}}$] 
  The values in parentheses are the arithmetic averages 
  over the best fits for the individual observational profiles for each case. 
  Please note that the errors in $\gamma$ and $\sigma$ are not independent 
  (refer to the text for more details).
  The value in front of the parentheses is the $\varepsilon$-weighted mean value 
  projected onto the mean MET. 
  \item[$^{\mathrm{c}}$] 
  The MET below the ``knee'' is approximated by  
  $\sigma_\mathrm{MET,b} = \sigma_C + (\Delta\sigma/\Delta\gamma) \cdot \gamma$. 
  The positions of the individual METs have a rms variation of  
  $\delta\gamma \sim 0\,\farcs001$ and $\delta\sigma \sim 0\,\farcs002$ - $0\,\farcs003$, 
  which is of the order of the grid increment. 
  Only the METs for the eclipse at 555\,nm show a larger variation with 
  $\delta\gamma \sim 0\,\farcs002$ and $\delta\sigma \sim 0\,\farcs005$. 
  \item[$^{\mathrm{d}}$]   The range of reasonable PSFs is limited by the 
  position just at the ``knee'' of the mean MET and on the other side by a 
  convolution with a pure Lorentzian 
  ($\sigma = 0$).
  \end{list}
\end{table*}

\subsection{Non-ideal PSFs incl. convolution with a Voigt function} 
\label{sec:artlimbnonideal}

Now those PSFs are used that are calculated by convolution of the 
diffraction-limited PSFs and a Voigt function. 
Synthetic intensity profiles are again derived from degraded images of the 
artificial eclipse and the Mercury transit and are then compared to the observed 
intensity profiles. 
An example for the Mercury transit is displayed in Fig~\ref{fig:artobs}.

A constant intensity offset $I_\mathrm{o}$, which is added to the synthetic profiles 
$I_\mathrm{syn,org}$, is introduced as a third parameter:  
\begin{equation}
	I_\mathrm{syn}\,(x, \gamma, \sigma, I_\mathrm{o}) = 
	I_\mathrm{o} + (1-I_\mathrm{o})\,I_\mathrm{syn,org}\,(x, \gamma, \sigma)
\end{equation}
The discrepancy between observed and synthetic profiles is measured for all 
parameter combinations $[\gamma, \sigma, I_\mathrm{o}]$. 
The quality of a fit is evaluated with the quantity 
\begin{equation}
	\label{eq:psffitgood}
	\varepsilon\,(\gamma, \sigma, I_\mathrm{o}) = 
	\left\langle \left( 
	\frac{I_\mathrm{obs} (x) - I_\mathrm{syn}\,(x, \gamma, \sigma, I_\mathrm{o})}{I_\mathrm{syn}\,(x, \gamma, \sigma, I_\mathrm{o})}
	\right)^2 \right\rangle_x
\end{equation}
averaged over the terminator region and the first 10\,\arcsec of the dark part in 
case of an eclipse and over the first 5\,\arcsec in case of the Mercury transit), 
respectively. 
This goodness-of-fit measure is in principle a sum-of-squares or least squares 
approach.
For every $[\gamma,\sigma]$ the fitting procedure returns an error vector for the 
different $I_\mathrm{o}$. 
Although the general picture remains the same when choosing a fixed 
$I_\mathrm{o}$, the minimum error is chosen for each $[\gamma,\sigma]$, 
effectively reducing the measured errors to $\varepsilon_\mathrm{min} (\gamma,\sigma)$.
The procedure is repeated for every observed profile.  

\paragraph{Reference case.}
First a reference case is constructed in order to investigate the fitting routine and 
the quality of the selected error measure. 
Instead of fitting an observed intensity profile, a known synthetic profile for 
$\gamma = 0\,\farcs005$ and $\sigma = 0\,\farcs01$ is preset. 
The error $\varepsilon_\mathrm{min}$ forms a distinct pattern in the $\gamma$-$\sigma$ 
plane, which is shown in Fig.~\ref{fig:psffitref}. 
The black track, where the error is smallest, is called here the ``minimum error  
trench'' (MET).
The properties of the Voigt function make the MET bend and form a ``knee''. 
In the reference case depicted in Fig.~\ref{fig:psffitref} the ``knee'' is located 
at $\gamma \approx 0\,\farcs03$ and $\sigma \approx 0\,\farcs016$. 
Below the knee, i.e. for smaller $\sigma$, the MET follows lines of constant Strehl 
but also roughly lines of equal broadening of the FWHM of the PSF core with 
respect to the pure diffraction-limited PSF. 
In contrast the part of the MET above the knee follows a line of constant 
damping parameter $a$ of the Voigt function (see Eq.~(\ref{eq:voigtdamp})). 
The MET in the reference case fades again along the upper right part, indicating 
that the fit between the intensity profile degenerates for increasing $\sigma$. 
The determined best-fitting parameter pair (marked with a circle in the plot) indeed 
matches the preselected one, demonstrating that the method in principle works.  

\begin{figure*}[t]
  \centering
  \includegraphics{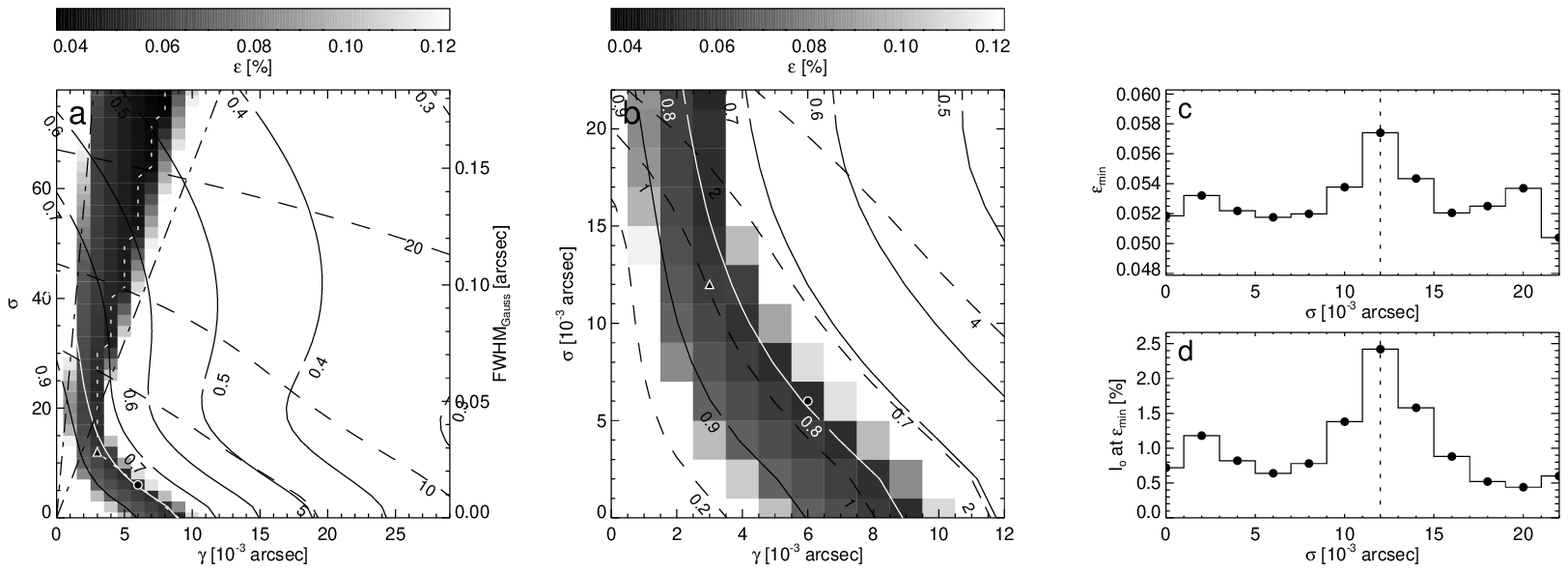}
   \caption{Goodness $\varepsilon$ of the fits between artificial 
    and observed intensity profiles for a selected image of the Mercury transit at 
   	$\lambda = 555$\,nm.   
	The synthetic profiles are generated with non-ideal PSF that are calculated 
	by convolution with a Voigt function. 
	See Fig.~\ref{fig:psffitref} for a detailed explanation of the figure. 
      }
   \label{fig:psffitcvo}
\end{figure*}

\paragraph{Fitting observed profiles.}
\label{sec:psffit_cvo}
The same procedure is now used for fitting the observed intensity profiles 
(see Sect.~\ref{sec:obsprofiles}).
Although the MET is also clearly pronounced in Fig.~\ref{fig:psffitcvo}, the best 
fit (not shown) is found just at the upper limit of the considered parameter 
range for $\sigma$. 
This parameter combination is clearly unreasonable as it implies a Strehl ratio 
of less than 0.5 and also a FWHM of the PSF core that would be 30\,\% 
broader than for the diffraction-limited PSF. 
It has to be attributed to imperfections of the fits due to anisotropic
stray-light contributions (see Sect.~\ref{sec:discus}).
Obviously additional constraints are required. 
The spatial power spectral density of granulation images observed with SOT 
would in principle allow to estimate the effective spatial resolution of 
the instrument. 
It would correspond to a line of constant FWHM or broadening in the 
$\gamma$-$\sigma$-plane (e.g., a dashed line in Fig.~\ref{fig:psffitcvo}), 
which would restrict the relevant MET and the corresponding parameter combinations 
to the part below. 
Preliminary tests show, however, that it is difficult to determine a strict limit 
for the spatial resolution. 
Here the Strehl ratio is used as additional constraint. 
Based on pre-flight measurements, \citet{tsuneta_2008_SOToverview} and 
\citet{2008SoPh..tmp...26S} state that the individual Strehl ratios of the 
optical telescope assembly (OTA) and the focal-plane package (FPP) both are 
0.9 or better at a wavelength of 500\,nm and that the combined Strehl ratio is 
thus certainly 0.8 or better (see \citeauthor{tarbell08fpp}, in prep., for 
details on FPP). 
Nevertheless it is necessary to test the optical performance under on-orbit 
conditions.
Indeed the MET below the ``knee'' follows a contour of a Strehl ratio close to 
$0.8$ for the Mercury transit at 555\,nm (see Fig.~\ref{fig:psffitcvo}) and 
668\,nm, whereas it is close to 0.85 for 450\,nm (see linear regression section 
of Table~\ref{tab:psffit}. 
The METes for the eclipse are very similar and favour Strehl ratios of 0.84 
to 0.86.

In the following the MET above the ``knee'' is neglected because of the too small  
Strehl ratios. 
Unfortunately the best parameter set [$\gamma, \sigma$] is less obvious than in 
the reference case. 
Nevertheless Fig.~\ref{fig:psffitcvo}c points at $\gamma = 0\,\farcs006$ and 
$\sigma = 0\,\farcs006$ as the parameter pair that produces the best fit 
for that particular Mercury transit image. 
For the other cases the best pair is found in different locations along the 
MET below the ``knee'' and also the exact position of the MET varies slightly. 
Obviously it is difficult to pin down a definite PSF that can reproduce all 
observations. 
On the other hand the resulting parameter range represents PSFs that are very 
similar in terms of FWHM and Strehl ratio. 
The arithmetic averages and standard deviations of $\gamma$ and  $\sigma$ 
are listed in Table~\ref{tab:psffit} for all three continuum bands for the 
Mercury transit and the total eclipse. 
The variation in $\sigma$ can be quite significant, e.g. in the case of the total 
eclipse at 555\,nm. 
The variations of $\gamma$ and  $\sigma$ are not independent and should rather 
be replaced by the deviation from the mean MET. 
As the MET below the ``knee'' follows roughly lines of constant Strehl ratio 
and broadening, the corresponding PSFs are very similar. 
This part of the MET can be approximated with a linear regression that accounts 
for all fits of the individual observations (see Table~\ref{tab:psffit}. 
The average and standard deviation of the slope $\Delta\sigma/\Delta\gamma$
and the offset $\sigma_C$ provide a range of METs, which can be translated into 
a meaningful error margin in $\gamma$ and $\sigma$.  
It is $\delta\gamma \sim 0\,\farcs001$ and 
$\delta\sigma \sim 0\,\farcs002$ - $0\,\farcs003$, 
which is of the order of the grid increment. 
Owing to the larger variation in the observational intensity profiles for the 
eclipse, the METs for the eclipse at 555\,nm vary by  
$\delta\gamma \sim 0\,\farcs002$ and $\delta\sigma \sim 0\,\farcs005$. 

The optimum PSF for each case can now be determined from the 
$\varepsilon$-weighted mean value projected onto the mean MET. 
The resulting [$\gamma, \sigma$] is essentially identical with 
the arithmetic averages or at most deviates in $\gamma$ by just of the order of 
the grid increment. 
For the Mercury transit a $\sigma_\mathrm{opt} = 0\,\farcs008$ is found for all 
three wavelengths, whereas $\gamma_\mathrm{opt}$ increases with wavelength from 
$0\,\farcs004$ to $0\,\farcs006$. 
The results for the green and red continuum are very similar in Strehl ratio $S$ (0.78-0.79)
and broadening $b$ of the central PSF lobe (1.4\,\% - 1.3\,\%).
In contrast the blue continuum indicates a slightly better instrument performance 
with $S = 0.84$ and $b = 0.8$\,\%.
The intensity offset for the optimum PSFs is derived from the average of the 
individual $\gamma$-$\sigma$-planes over all individual observations for each case
at the parameter position [$\gamma_\mathrm{opt}, \sigma_\mathrm{opt}$]. 
It is only 0.3\,\% to 0.4\,\% of the mean solar disc intensity. 

The eclipse case indicates an excellent instrument performance very close to the 
blue continuum Mercury transit case. 
On the other hand larger intensity offsets of 2.5\,\% and 3.0\,\% are necessary 
for the green and blue continuum, respectively, clearly demonstrating the larger 
uncertainties arising from the variation among the eclipse profiles.  
In addition there could be systematic differences between the Mercury transit 
and the eclipse observations due to the different observation dates. 
Some instrumental properties change with time under space conditions, 
causing for instance a gradual shift of 
the focus position \citep{2008arXiv0804.3248I}. 

Finally the range of reasonable PSFs can be limited by the cases 
just at the ``knee'' of the mean MET and on the other side by a convolution with 
a pure Lorentzian ($\sigma = 0$, see Sect.~\ref{sec:psffit_clo}).
The indicated instrument performance is very close to those of the optimum PSFs.  
See Table~\ref{tab:psffit} for details. 

\begin{figure}[t]
  \includegraphics{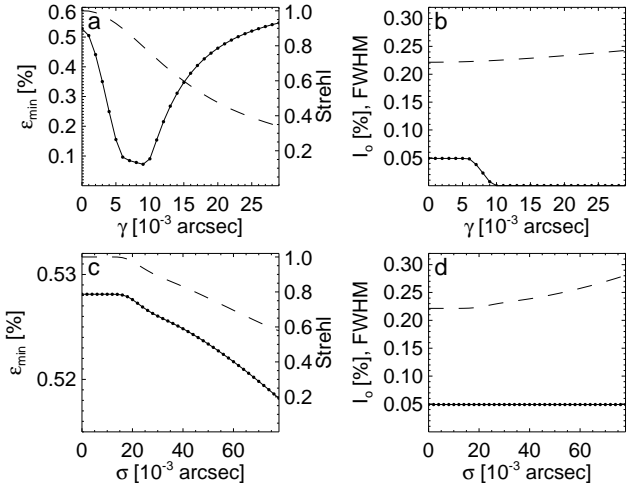}
   \caption{Goodness $\varepsilon$ (left column, solid lines) of the fits between 
   	artificial and observed intensity profiles for the Mercury transit at 
   	$\lambda = 555$\,nm and applied intensity offset $I_\mathrm{o}$ 
	(right column, solid lines) for PSFs with convolution with a Lorentzian 
	(top row) and a Gaussian (bottom row).   
	The goodness is averaged over the individual observations. 
	The dashed lines are the Strehl ratio in the left column and the FWHM of 
	the combined PSF in the right column, respectively.
      }
   \label{fig:psffitclg}
\end{figure}
\subsection{PSFs including convolution with a Lorentzian}
\label{sec:psffit_clo}

For the case $\sigma = 0$ the deviation from the diffraction-limit essentially 
reduces to a pure Lorentzian with a FWHM of $\gamma$. 
Fitting the observed profiles requires values for $\gamma$ between 
0\,\farcs007 and 0\,\farcs009 (see Table~\ref{tab:psffit} and 
Fig.~\ref{fig:psffitclg}a, cf. Fig.~\ref{fig:psffitclg}a). 
For the green and red continuum Mercury transit case the best fits imply PSFs 
with a Strehl ratio of $S = 0.79$ and a broadening of the central lobe 
with respect to the diffraction-limited PSF by 1.3\,\% and 1.2\,\%, respectively. 
The other cases favour Strehl ratios of 0.84 to 0.87 and  a broadening of 0.7\,\%
to 0.8\,\%. 
The intensity offset $I_\mathrm{o}$ is in most cases somewhat higher than for 
the optimum PSFs for a convolution with a Voigt function (see Sect.~\ref{sec:psffit_cvo})
but stays of the order of 1\,\% for the Mercury transit and up to 
$3.2 \pm 0.4$ for the red continuum eclipse.   

\begin{figure*}[t]
   \sidecaption
   \includegraphics{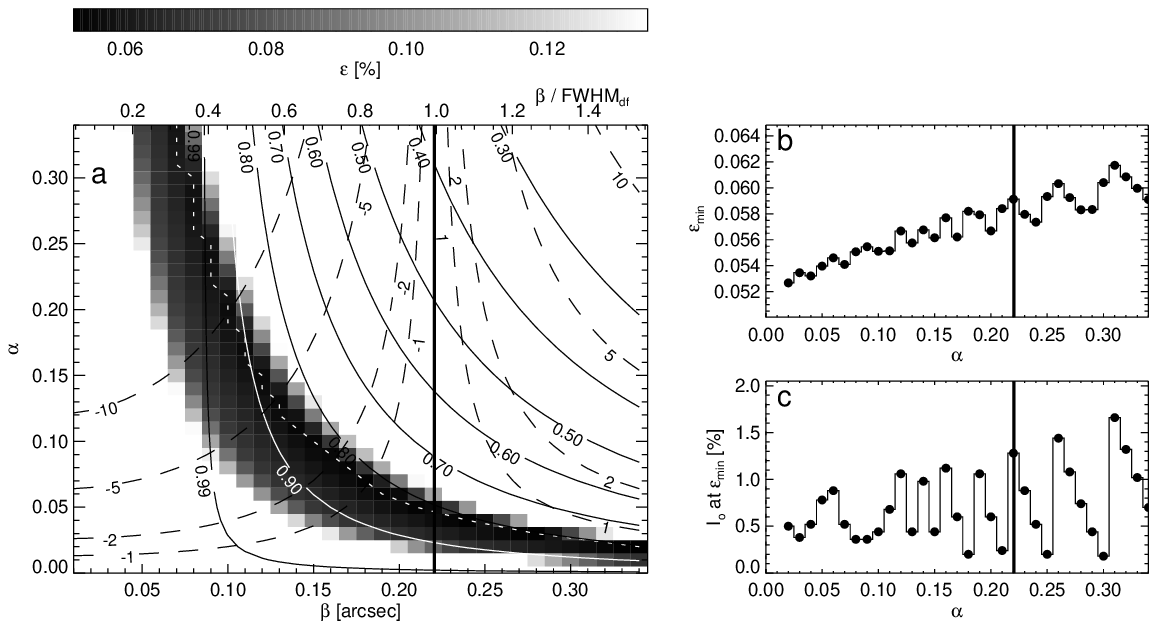}
   \caption{Goodness $\varepsilon$ of the fits between artificial 
    and observed intensity profiles for the Mercury transit at 
   	$\lambda = 555$\,nm (cf.~Fig.~\ref{fig:psffitref}). 
	The non-ideal PSF contribution are calculated by linear combination with a 
	\textbf{a)} The goodness $\varepsilon$ as defined in 
	Eq.~(\ref{eq:psffitgood}) as function of $\alpha$ and $\beta$, 
	\textbf{b)} $\varepsilon$ along the ``minimum error trench'' (MET, marked 
	with white dotted line in panel~a), and
	\textbf{c)} the applied intensity offset $I_\mathrm{o}$ along the MET.  
	Overplotted in panel a are contours of constant Strehl ratio 
	(solid) and of constant broadening (dashed) of the combined PSF.
	The broadening is given in percent points increase of the FWHM of the PSF
	($\mathrm{FWHM}_\mathrm{PSF}$) with  
	respect to the one of the diffraction-limited PSF 	($\mathrm{FWHM}_\mathrm{dl}$).
	All parameter combinations with a $\beta$ smaller than $\mathrm{FWHM}_\mathrm{dl}$
	(with except for combinations with \mbox{$\alpha = 0$)} 
	produce unphysical PSFs with 
	$\mathrm{FWHM}_\mathrm{PSF} < \mathrm{FWHM}_\mathrm{dl}$. 
	The cases are found left of the thick solid line, which marks 
	$\mathrm{FWHM}_\mathrm{PSF} = \mathrm{FWHM}_\mathrm{dl}$ (in all panels).}
   \label{fig:psffitllo}
\end{figure*}

\subsection{PSFs including convolution with a Gaussian}
\label{sec:psffit_cga}

Convolution of a diffraction-limited PSF with only a pure Gaussian does produce
fits to the observed intensity profiles that are usually worse than for the 
analogeous cases with a Lorentzian (Sect.~\ref{sec:psffit_clo}) or better a 
Voigt function (Sect.~\ref{sec:psffit_cvo}). 
The goodness $\varepsilon$ displayed in Fig.~\ref{fig:psffitclg}c does not 
indicate any preferred parameter $\sigma$. 
Obviously a pure Gaussian is generally not a suitable approximation for the 
non-ideal contributions to a PSF, especially for its wings. 
Although both the convolution with a Gaussian and Lorentzian lead to an increase 
in FWHM and Strehl ratio with their primary parameter ($\sigma$ and $\gamma$, 
respectively), the details of the resulting PSF profiles differ. 
Already for small values of $\sigma$ the Gaussian causes the central lobe of the 
combined PSF to get so broad that the first minimum is removed and the central 
and secondary lobes essentially fuse.  
The higher order lobes are smoothed out equally fast. 
The Lorentzian obviously has the potential to produce profiles that are closer 
to reality. 

\subsection{Application of non-ideal linearly combined PSFs}

Despite conceptual concerns the individual observed profiles can be fitted 
reasonably well when using a linear combination of an ideal PSF 
$\mathcal{P}_\mathrm{dl}$ and an additional Lorentzian (see Eq.~\ref{eq:psflor} 
and Fig.~\ref{fig:sotpsf}).
Owing to the large fluctuation among the observed intensity profiles (see 
Fig.~\ref{fig:sotstray}), the best combination of $\alpha$ and $\beta$, 
however, varies strongly. 
For instance, the observed eclipse profiles at 555\,nm are best matched with 
the artificial eclipse experiments when adding a Lorentzian with 
\mbox{$\alpha =  0.17 \pm 0.10$} and 
\mbox{$\beta  =  0.07 \pm 0.03$}, while a residuum of 
\mbox{$\Delta I_\mathrm{res} =  (2.9 \pm  1.3)$\,\%} remains. 
But the range of typical parameters is extended: 
\mbox{$\alpha =  0.05 - 0.30$} and 
\mbox{$\beta  =  0.05 - 0.10$} with 
$\Delta I_\mathrm{res} =  (0.5 - 4.2)$\,\%.  
Similar values are found for the eclipse at 555\,nm and 668\,nm but also for 
the Mercury transit at 450\,nm. 
There are individual cases for which $\alpha$ can be up to 0.35 but an 
increase in $\alpha$ is compensated by a relatively small value in $\beta$ 
and vice versa. 
In analogy to the convolution approach in Sect.~\ref{sec:psffit_cvo} the 
goodness of fit shows this clear relation between $\alpha$ and $\beta$ (see 
Fig.~\ref{fig:psffitllo}).
Obviously there is a certain ambiguity in the parameter pair $[\alpha, \beta]$. 
For the other Mercury transit cases at 555\,nm and 668\,nm rather values around 
\mbox{$\alpha =  0.05 - 0.10$} and 
\mbox{$\beta  =  0.15 - 0.25$} with 
\mbox{$I_\mathrm{res} =  (0.2 - 1.0)$\,\%} prevail. 
Considering all profiles for all wavelengths and Mercury transit and eclipse 
together the following value ranges are found: 
\mbox{$\alpha =  0.15$ (0.05 - 0.30),} 
\mbox{$\beta  =  0.10$ (0.05 - 0.25),} 
\mbox{$(I_\mathrm{res} =  1.8$\,\% (0.0 - 4.5)\,\%.} 
The most frequent combination of $[\alpha, \beta]$ is $[0.25, 0.05]$ 
with 17\,\% of all cases, followed by $[0.20, 0.05]$ and $[0.05, 0.25]$ 
(both 13\,\%).

As this approach suffers from an ambiguity between $\alpha$ and $\beta$, 
additional constraints are necessary -- just like the convolution approach in 
Sect.~\ref{sec:psffit_cvo}.  
On the other hand some cases have to be discarded because they are unphysical, 
e.g. when the Strehl ratio would be greater than 1.  
In general $\beta$ should be equal to or larger than the FWHM of the 
diffraction-limited PSF. 
Hence the parameter domain left of the thick solid line in Fig.~\ref{fig:psffitllo}a, 
which marks $\beta = \mathrm{FWHM_{dl}}$, is invalid.

\section{Discussion}
\label{sec:discus}

\paragraph{Suitable observations.}
Observations of the solar limb have fundamental problems with respect to the 
determination of a PSF. 
The centre-to-limb variation induces an intensity gradient across the image, 
which makes a limb observation an inhomogenously illuminated and thus non-ideal 
test case. 
The corresponding stray-light level might be strongly underestimated as the overall 
light level is much smaller than at disc-centre (see Fig.~\ref{fig:sotcnt}).
In addition the centre-to-limb variation complicates the definition of a reference 
intensity for the stray-light level. 

In general it is preferable to have a homogeneous background with a well-defined 
proxy for the PSF. 
Ideal is a case which is not too different from regular observations. 
In this respect a Mercury transit seems to be the best available test case. 
The overall light level in the telescope is not significantly affected if the 
the FOV is much larger than the occulting Mercury disc.  
The radial geometry is very convenient as radial averaging offers a way to 
deal with the anisotropy of the PSF and to describe it with an average intensity 
offset instead. 
A clear drawback of course is that a Mercury transit 
does only provide the first 5\,\arcsec of the PSF profile. 

In contrast even the far wings can be analysed for an eclipse image. 
On the other hand eclipses have a clear disadvantage if there are 
anisotropic contributions that vary across the FOV. 
In that case, as found here for SOT, the exact position of a eclipse 
terminator in the image introduces another degree of freedom. 
Also the larger relative occulted image area, which varies from image to image, 
make eclipse image depend much stronger on the overall light level in the telescope. 

Ideally one should try to combine the analysis of both, Mercury transits and 
eclipses, whenever possible, where images away from the solar limb are preferable. 

\paragraph{PSF models.}
It is not obvious which mathematical model is best for representing 
the non-ideal contributions of a PSF. 
Already \citet{1984ssdp.conf..174N} made clear that PSFs of comparable quality 
can be constructed from a combination of two Lorentzians instead of Gaussians. 
Another approach is used by \citet{deForest_TRACE_PSF}, who model the PSF of the 
Transition Region and Coronal Explorer (TRACE) with the sum of a narrow Gaussian 
core and Lorentzian wings. 

An important finding is that it is indeed crucial to use realistic detailed PSFs. 
Ideally one would assemble a realistic PSF by following the optical path step by 
step and adding a component for each optical element.
Unfortunately this is not possible in most cases as it requires detailed 
measurements and testing of each element under conditions that are very close to 
those of the observations. 
In a real optical system chromatic aberration results in a systematic change 
of the optical properties with wavelength. 
In case of SOT, the chromatic aberration is small for the OTA 
\citep{2008SoPh..tmp...26S} and also for the BFI and can thus usually be neglected. 
An exception may be simultaneous observations that span a large wavelength range, 
resulting in a still small but possibly significant effect \citep{2008arXiv0804.3248I}. 
Even an interchange of intensity between different wavelengths is in principle 
possible.  
A PSF for a certain wavelength thus could gain or lose energy from other 
wavelengths.
Here the PSFs are treated as purely monochromatic because the aforementioned 
effects are considered negligible in comparison to other uncertainties. 

Simplifications are actually unavoidable but one should at least account for 
instrumental details such as spiders and a central obstruction, as it is the case 
for SOT. 
In particular the spider legs do not only block the telescope pupil but  also 
scatter the incoming light at small angles that will inevitably broaden the PSF. 
The strong variation in eclipse profiles and the number of potential causes, 
however, limit the number of meaningful free model parameters. 
As long as an approximation produces acceptable fits to the intensity profiles, 
it should thus be as simple as possible. 
For instance, the net effect of a possible defocus is in principle included in 
the PSFs determined here although it cannot explicitly be separated from other 
effects.
The introduction of a defocus term would certainly complicate the model without 
producing better fits of the intensity profiles. 

A linear combination of a diffraction-limited PSF and a Lorentzian should only be 
used for sufficiently large $\beta$, which is mostly the case when the deviations 
are dominated by the seeing induced by the Earth's atmosphere
\citep[see, e.g.][]{2007ApJ...655..615L, leenaarts05}. 
For space-borne instruments smaller values of $\beta$ and thus narrower 
Lorentzians can occur, as seen here for example of SOT. 
For very small $\beta$ the FWHM of the combined PSF gets smaller than the one for 
the diffration-limited PSF, whereas the Strehl ratio can exceed~1. 
In those cases a linear combination is obviously unphysical. 
A Gaussian as PSF, although often used for the sake of simplicity and/or the lack 
of constraints on the PSF, is certainly not a good approximation, as it grossly 
underestimates the wing contributions.
Instead one should whenever possible prefer a convolution of a detailed 
ideal PSF and a Voigt function or possibly a more realistic description of 
the non-ideal contributions. 

\paragraph{Limitations of the detailed PSFs for SOT.}

Anisotropic contributions and the complex dependence on the overall light level 
in the optical system and on instrumental details such as focus position render it 
simply impossible to determine only one PSF that is suitable for all situations. 
It is certainly advisable to instead use a range of detailed PSFs, which -- as 
discussed in Sect.~\ref{sec:artlimbnonideal} for the Voigt-convolution approach 
-- is limited by the extreme PSFs at the MET knee and for a pure Lorentzian
(see Table~\ref{tab:psffit}). 
Both are still similar in terms of Strehl ratio and FWHM of the central lobe. 
Applying these PSFs in addition to the best-fit PSFs provide an error margin for 
the quantity under investigation.  

The PSFs constructed here combine the instrumental effects of the optical 
telescope assembly (OTA) and the Broadband Filter Imager (BFI).  
It is not clear to what extent the results can be applied to observations with the 
Narrowband Filtergram Imager (NFI) and the spectro-polarimeter (SP) because the 
last part of the optical path inside the focal plane package differs. 
In principle the PSF determination should be repeated for the other instruments 
but no suitable observations of the Mercury transit were found. 

A remaining uncertainty concerns the normalisation of the PSF to its integral, 
to which the far wings in principle contribute. 
The surface integrals of the Voigt functions diverge in the limit of infinite 
angles as the Lorentzian contribution converges too slowly. 
Therefore a Lorentzian needs to be truncated in order to keep the total energy of 
the PSF finite. 
Here it is limited by the array size of the combined PSF. 
From a practical point of view one can argue that the integrals of the PSFs in 
good approximation approach an asymptotic value because the growth in area is more 
than compensated with the steep decrease of the PSF profile.
The use of a (truncated) Voigt function is still justified considering that in view 
of the many sources of uncertainty it is only an approximation after all.  
\citet{deForest_TRACE_PSF} truncate the non-ideal contributions to the PSF 
wings for TRACE by applying a Gaussian envelope.  
Nevertheless a proper normalisation of the PSF requires a sufficiently large 
array (here a few 10\arcsec in each direction). 
As a consequence the exact Strehl ratio depends on the extent of the PSF.  
Nevertheless the Strehl ratios of the PSFs constructed here are of the order 
of 0.8 or better and thus agree with \citet{tsuneta_2008_SOToverview}. 

\paragraph{Stray-light estimate for SOT.}
Neither can one give a single number for the stray-light contribution but at best 
a rough estimate for the upper limit of the stray-light contribution. 
As such one can use the difference between the observed residual intensity and the 
contribution due to the wings of the ideal (theoretical) PSF. 
The artificial Mercury transits for the best-fitting PSFs (see Table~\ref{tab:psffit}) 
have intensities of 6.5\,\%, 10.0\,\%, and 10.8\,\% at a distance of 5\arcsec\ from 
the terminator for green, blue, and red continuum, respectively. 
These values are close to the estimated residual intensity in ``regular'' granulation 
images (Table~\ref{tab:iresest}). 
The small Mercury disc affects the overall light level only slightly, making it a 
good test case.
Subtracting the intensities at 5\arcsec\ derived with the diffraction-limited PSF 
(see Table~\ref{tab:dfpsfires}) yields 4.7\,\%, 7.8\,\%, and 8.2\,\%, which can be 
considered as approximate upper limit for the stray-light contribution at that distance 
for high light levels.  
The value for the green continuum is significantly higher than judging from the eclipse 
image in Fig.~\ref{fig:sotstrayimg} but the latter of course has a lower overall 
light level, which causes the residual intensities to be smaller (see 
Fig.~\ref{fig:sotcnt}).

\paragraph{Fitting procedure.} 
Fitting a profile across a terminator is difficult as it both has an extremely steep 
slope and a long tail at low intensities. 
Depending on the chosen measure either one or the other is weighted more. 
As it is not obvious which approach is the best, the least squares method chosen here 
is kept as simple as possible.  
That the choice is reasonable is supported by the fact that the resulting MET in the 
$\gamma$-$\sigma$-plane indeed coincides with the pre-flight measurements of the Strehl 
ratio \citep{tsuneta_2008_SOToverview}. 
Removing the denominator in Eq.~(\ref{eq:psffitgood}) gives more weight to the 
difference between observed and synthetic profile directly at the steep slope. 
The result is a small shift of the MET in the $\gamma$-$\sigma$-plane. 
For the Mercury transit the MET is shifted upwards by 
$\Delta\sigma = 0\,\farcs001-0\,\farcs002$ and to the right by 
$\Delta\gamma \leq 0\,\farcs001$, which is just of the order of the resolution 
of the $\gamma$-$\sigma$-plane. 
The Strehl ratio along the MET below the ``knee'' increases by just 0.01 to 0.03. 
The quality measure given in Eq.~(\ref{eq:psffitgood}) produces balanced fits, 
whereas removing the denominator can cause significant deviations in the tail.  
The eclipse is slightly more susceptible to changes of the quality measure 
$\varepsilon$, producing shifts of the MET of 
$\Delta\gamma = 0\,\farcs001-0\,\farcs004$ and 
$\Delta\sigma = 0\,\farcs002-0\,\farcs008$ and a change of the Strehl ratio of 
up to $\sim 0.10$ in some cases. 
Reducing the spatial range that is considered for $\varepsilon$ from 10\arcsec\ to 
5\arcsec\ effects the position of the MET much less with shifts comparable to the 
above-mentioned uncertainties for the Mercury transit.  

In summary it can be concluded that the Mercury transit case is less 
susceptible to details of the fitting procedure than the eclipse case. 

\paragraph{Intensity offset.}
Although the non-ideal PSFs provide a much better fit of the observed intensity 
profiles than the ideal ones small quantitative discrepancies remain. 
The remaining differences reflect the fact that the detailed PSF models 
constructed here are still a necessary simplification whereas the real PSFs 
additionally include anisotropy of the residual light and thus depend on the 
position in the FOV. 
The latter would require more constraints, which are usually -- as in the present 
case -- not available. 
The introduction of the constant intensity offset $I_\mathrm{o}$ offers a way of 
capturing the net effect of the position-dependence and improves the fit of the 
intensity profiles. 
Furthermore $I_\mathrm{o}$  absorbs uncertainties in the determination of reference 
intensities and of the fitting procedure but also potential small contributions 
from the very far PSF wings. 
The values for $I_\mathrm{o}$, however, are mostly only of the order of 1\,\% of 
the mean solar disc intensity or less for the Mercury transit (see, e.g., 
Fig.~\ref{fig:psffitcvo}c). 
The variation of the mean count rate and the anisotropy of the stray-light 
produce a larger remaining uncertainty, requiring intensity offsets of up to 
$\sim 3.6$\,\% (for the optimum PSFs with convolution of a Voigt function).

\paragraph{Anisotropic contributions.}
For an observed profile anisotropic contributions due to instrumental scattered 
light -- although small -- are expected and indeed found in case of SOT  
(Fig.~\ref{fig:sotstrayimg}).
Adding an intensity contribution of the form $I_\mathrm{add} (x) = x\,\frac{d}{dx}I$ 
to the reference case changes the slope of the synthetic intensity profiles across 
the terminator and removes the otherwise clear fading of the MET on the upper 
branch. 
The parameter pair producing the best fit, i.e. the minimum error, in the 
$\gamma$-$\sigma$ plane moves along the MET to the upper right to Strehl ratios 
that are certainly unreasonable. 
In reality the slope changes are certainly more complex than the simple linear 
contribution tested here. 
Consequently the used synthetic intensity profiles cannot perfectly match the 
observed profiles. 
There will always be a remaining (small) discrepancy. 
An exact match would require a detailed model of the anisotropic contributions, 
which would demand for more observational constraints on the PSF properties.  
Obviously anisotropic contributions severly complicate the determination of a 
reasonable PSF. 

\paragraph{Alternative methods.}
The direct determination of a (two-dimensional) PSF from an observational image of 
an eclipse would in principle be superior to the forward approach presented here. 
In most cases, however, a direct determination is rather hopeless in view of 
uncertainties that arise from the PSF wings but also from detector sensitivity, 
data sampling and many more. 
All these effects make it difficult to retrieve a definite analytical fit of 
a measured intensity profile. 
While a radially symmetric PSF can be derived this way but asymmetric contributions 
due to, e.g., instrumental stray-light are usually hard to determine.  
Obviously non-radial features such as the imprint of spider legs cannot be 
captured this way. 

One has to keep in mind that the situation is quite different from night time 
observations for which stars, i.e. point sources, offer an easy way to determine 
a PSF.
It may be possible to directly measure a two-dimensional PSF for a solar telescope 
by taking images of background stars far away from the solar disc but there the 
conditions are fundamentally different from those of regular solar observations. 
As the dependence on the overall light level in the telescope effects in particular 
the PSF wings, a measurement outside the solar disc may produce a PSF that cannot 
be applied to observations of the bright solar disc. 
A potential technical problem is that solar telescopes and in particular the detectors 
are usually not designed for such low-intensity situations so that the low signal-to-noise 
ratio prevents the direct measurement of a PSF. 

The analysis of very small features in principle provides an indicator of the effective 
spatial resolution and with it the FWHM of the PSF. 
\citet{2008SoPh..tmp...26S} use a intensity profile of a small G-band bright point 
and conclude that the OTA of SOT indeed operates close to the diffraction-limit
\citep[see also][]{2008arXiv0804.3248I}.  
Unfortunately this method does not yield further information about the detailed 
structure of the PSF. 
The same is true for the spatial power spectral density of quiet Sun granulation 
images.

Another common method is to determine a PSF by matching the intensity distribution 
based on numerical simulations of the solar atmosphere with observations.  
This approach should only be considered in the total absence of other independent 
methods as it assumes the numerical simulations to be perfectly realistic. 

\section{Conclusion}
\label{sec:conc}

The analysis of Mercury transit and eclipse broadband filtergrams shows that SOT 
onboard the Hinode satellite is an excellent instrument, which performs close to 
the diffraction limit. 
In this work PSFs are constructed by a convolution of the ideal diffraction-limited 
PSF with a narrow Voigt function, which accounts for non-ideal contributions due to 
instrumental stray-light. 
The comparison of synthetic and observed intensity profiles across the Mercury 
and lunar terminator does unfortunately not produce a unique solution for the 
best-fitting PSF, which can account for all observational conditions.
Likely reasons are the anisotropy of the residual light and the dependence on the 
overall light level.
A realistic PSF would even depend on the position on the solar disc, owing to the 
change of the overall light level in the telescope due to the centre-to-limb variation
of the continuum intensity.  
Obviously it is impossible to provide a stray-light estimate as just a single 
(wavelength-dependent) number. 

Instead of a single PSF for each wavelength one should rather use a range of PSFs, 
which allow to translate the uncertainty in PSF properties into an error margin for 
the quantity under investigation. 
Fortunately the differences between synthetic and observed intensity profiles form 
a distinct relation of the Voigt parameters $\gamma$ and $\sigma$, which is named 
here the ``minimum error trench'' (MET). 
The empirically estimated Strehl ratio of the optical system serves as additional 
constraint, limiting the range of reasonable PSFs to a part of the MET where the 
PSFs are very similar in terms of Strehl ratio, FWHM of the central lobe, and thus 
their effect on the intensity distribution. 
There are remaining uncertainties due to the anisotropy and light-level-dependence 
of the residual intensity, which result in an intensity offset during the profile 
fitting process. 
For the Mercury transit the offsets are usually only of the order of 1\,\% of the 
solar disc intensity or less, whereas the eclipse images cause values of up 
3.6\,\% with a pronounced variation among the individual images. 
Obviously the Mercury transit is the more reliable test case. 

Nevertheless reasonable PSF estimates can be constructed (see Table~\ref{tab:psffit}).
They all require only narrow non-ideal contributions, indicating an excellent   
performance of the instrument.

\begin{acknowledgements}
  The author thanks M.~Carlsson, S.~Tsuneta, F.~W{\"o}ger, Y.~Suematsu, 
  L.~Rouppe van der Voort,  
  {\O}.~{Langangen}, O.~von der L{\"u}he, and O.~Steiner for useful hints and 
  helpful discussions. 
  T.~Berger is thanked for early flat-fields and dark frames for the SOT data 
  reduction and also for helpful information regarding technical details of data 
  and instrument.
  S.~Haugan and T.~Fredvik are acknowledged for their support with the Hinode 
  data centre. 
  An ideal G-band PSF for SOT was kindly provided by Y.~Suematsu. 
  This work was supported by the Research Council of Norway, grant 
  170935/V30, and a Marie Curie Intra-European Fellowship of the European Commission 
  (6th Framework Programme, FP6-2005-Mobility-5, Proposal No. 042049).   
  Intensive use was made of the Hinode Science Data Centre Europe hosted 
  by the Institute of Theoretical Astrophysics of the University of Oslo, Norway.
  Hinode is a Japanese mission developed and launched by ISAS/JAXA, collaborating 
  with NAOJ as a domestic partner, NASA and STFC (UK) as international partners. 
  Scientific operation of the Hinode mission is conducted by the Hinode science team 
  organized at ISAS/JAXA. This team mainly consists of scientists from institutes in 
  the partner countries. Support for the post-launch operation is provided by JAXA 
  and NAOJ (Japan), STFC (U.K.), NASA, ESA, and NSC (Norway). 
  This research has made use of NASA's Astrophysics Data System.
\end{acknowledgements}
\bibliographystyle{aa}

\begin{thebibliography}{17}
\expandafter\ifx\csname natexlab\endcsname\relax\def\natexlab#1{#1}\fi

\bibitem[{{DeForest} \& {Wills-Davey}(submitted)}]{deForest_TRACE_PSF}
{DeForest}, C.~E. \& {Wills-Davey}, M.~J. submitted, \apj

\bibitem[{{Deubner} \& {Mattig}(1975)}]{1975A&A....45..167D}
{Deubner}, F.~L. \& {Mattig}, W. 1975, \aap, 45, 167

\bibitem[{{Freeland} {et~al.}(2000){Freeland}, {Bentley}, \&
  {Murdin}}]{2000eaa..bookE3390F}
{Freeland}, S., {Bentley}, R., \& {Murdin}, P. 2000, Encyclopedia of Astronomy
  and Astrophysics

\bibitem[{{Ichimoto} {et~al.}(2008){Ichimoto}, {Katsukawa}, {Tarbell}, {Shine},
  {Hoffmann}, {Berger}, {Cruz}, {Suematsu}, {Tsuneta}, {Shimizu}, \&
  {Lites}}]{2008arXiv0804.3248I}
{Ichimoto}, K., {Katsukawa}, Y., {Tarbell}, T., {et~al.} 2008, ArXiv e-prints,
  804

\bibitem[{{Kosugi} {et~al.}(2007){Kosugi}, {Matsuzaki}, {Sakao}, {Shimizu},
  {Sone}, {Tachikawa}, {Hashimoto}, {Minesugi}, {Ohnishi}, {Yamada}, {Tsuneta},
  {Hara}, {Ichimoto}, {Suematsu}, {Shimojo}, {Watanabe}, {Shimada}, {Davis},
  {Hill}, {Owens}, {Title}, {Culhane}, {Harra}, {Doschek}, \&
  {Golub}}]{2007SoPh..243....3K}
{Kosugi}, T., {Matsuzaki}, K., {Sakao}, T., {et~al.} 2007, \solphys, 243, 3

\bibitem[{{Langangen} {et~al.}(2007){Langangen}, {Carlsson}, {Rouppe van der
  Voort}, \& {Stein}}]{2007ApJ...655..615L}
{Langangen}, {\O}., {Carlsson}, M., {Rouppe van der Voort}, L., \& {Stein},
  R.~F. 2007, \apj, 655, 615

\bibitem[{{Langhans} \& {Schmidt}(2002)}]{2002A&A...382..312L}
{Langhans}, K. \& {Schmidt}, W. 2002, \aap, 382, 312

\bibitem[{{Leenaarts} \& {Wedemeyer-B{\"o}hm}(2005)}]{leenaarts05}
{Leenaarts}, J. \& {Wedemeyer-B{\"o}hm}, S. 2005, \aap, 431, 687

\bibitem[{{Levy}(1971)}]{1971A&A....14...15L}
{Levy}, M. 1971, \aap, 14, 15

\bibitem[{{Mattig}(1983)}]{1983SoPh...87..187M}
{Mattig}, W. 1983, \solphys, 87, 187

\bibitem[{{Nordlund}(1984)}]{1984ssdp.conf..174N}
{Nordlund}, A. 1984, in Small-Scale Dynamical Processes in Quiet Stellar
  Atmospheres, ed. S.~L. {Keil}, 174

\bibitem[{{Sch{\"u}ssler} {et~al.}(2003){Sch{\"u}ssler}, {Shelyag},
  {Berdyugina}, {V{\"o}gler}, \& {Solanki}}]{2003ApJ...597L.173S}
{Sch{\"u}ssler}, M., {Shelyag}, S., {Berdyugina}, S., {V{\"o}gler}, A., \&
  {Solanki}, S.~K. 2003, \apjl, 597, L173

\bibitem[{{Shimizu} {et~al.}(2007){Shimizu}, {Nagata}, {Tsuneta}, {Tarbell},
  {Edwards}, {Shine}, {Hoffmann}, {Thomas}, {Sour}, {Rehse}, {Ito},
  {Kashiwagi}, {Tabata}, {Kodeki}, {Nagase}, {Matsuzaki}, {Kobayashi},
  {Ichimoto}, \& {Suematsu}}]{2007SoPh..tmp..154S}
{Shimizu}, T., {Nagata}, S., {Tsuneta}, S., {et~al.} 2007, \solphys, 154

\bibitem[{{Suematsu}(2007)}]{suematsu07privcomm}
{Suematsu}, Y. 2007, priv. comm.

\bibitem[{{Suematsu} {et~al.}(2008){Suematsu}, {Tsuneta}, {Ichimoto},
  {Shimizu}, {Otsubo}, {Katsukawa}, {Nakagiri}, {Noguchi}, {Tamura}, {Kato},
  {Hara}, {Kubo}, {Mikami}, {Saito}, {Matsushita}, {Kawaguchi}, {Nakaoji},
  {Nagae}, {Shimada}, {Takeyama}, \& {Yamamuro}}]{2008SoPh..tmp...26S}
{Suematsu}, Y., {Tsuneta}, S., {Ichimoto}, K., {et~al.} 2008, \solphys, 26

\bibitem[{{Tarbell} {et~al.}(to be submitted)}]{tarbell08fpp}
{Tarbell}, T. et al., to be submitted, \solphys

\bibitem[{{Tsuneta} {et~al.}(2008){Tsuneta}, {Ichimoto}, {Katsukawa}, {Nagata},
  {Otsubo}, {Shimizu}, {Suematsu}, {Nakagiri}, {Noguchi}, {Tarbell}, {Title},
  {Shine}, {Rosenberg}, {Hoffmann}, {Jurcevich}, {Kushner}, {Levay}, {Lites},
  {Elmore}, {Matsushita}, {Kawaguchi}, {Saito}, {Mikami}, {Hill}, \&
  {Owens}}]{tsuneta_2008_SOToverview}
{Tsuneta}, S., {Ichimoto}, K., {Katsukawa}, Y., {et~al.} 2008, \solphys, online

\end{thebibliography}

\end{document}